\documentclass[pdflatex,sn-mathphys-num]{sn-jnl}% Math and Physical Sciences Numbered Reference Style
\usepackage{graphicx}          % 插入图片
\usepackage{multirow}%
\usepackage[utf8]{inputenc}
\usepackage[T1]{fontenc}
\usepackage{amsmath,amssymb,amsfonts}%
\usepackage[ruled,linesnumbered,lined]{algorithm2e}
\usepackage{geometry}
\usepackage{booktabs}
\usepackage{amsthm}%
\usepackage{mathrsfs}%
\usepackage[title]{appendix}%
\usepackage{xcolor}%
\usepackage{textcomp}%
\usepackage{manyfoot}%
\usepackage{booktabs}%
\usepackage{algpseudocode}%
\usepackage{listings}%
\usepackage{afterpage} % 将 caption 推迟到下一页输出。
\newgeometry{top=2cm, bottom=2cm, left=2cm, right=2cm}
\usepackage{setspace}%
\usepackage{titlesec}       % 导入 titlesec 宏包
\titleformat{\paragraph}[block]  % 改为 block 形状（标题单独成行）
  {\normalfont\normalsize\bfseries}  % 正体 + 正常字号 + 粗体
  {\theparagraph}                    % 显示段落编号（如果不需要可删除）
  {1em}                              % 编号与标题的间距
  {}                                 % 标题前无额外内容
\newcounter{edfig}% 独立的 Extended Data Figure 计数器
\renewcommand{\theedfig}{% 定义编号显示格式
Extended Data Fig.~\arabic{edfig}%
}
\theoremstyle{thmstyleone}%
\theoremstyle{thmstyletwo}%

\theoremstyle{thmstylethree}%

\usepackage{indentfirst}          % 让章节后的第一段也缩进
\usepackage{graphicx}
\usepackage{caption} % 1. 引入宏包

\begin{document}

\title[Article Title]{Imaging cellular-level brain microstructure with diffusion MRI}


% ============ 作者信息 ============

\author[1]{\fnm{Xiaodong} \sur{Li}}

\author[2]{\fnm{Jing} \sur{Zhao}}
\author[2]{\fnm{Baolan} \sur{Lu}}
\author[2]{\fnm{Jinzhu} \sur{Wang}}

\author[3]{\fnm{Xinhua} \sur{Wei}}

\author[4]{\fnm{Qingxian} \sur{Yang}}

% ============ 单位信息 ============

\affil[1]{\orgdiv{Laboratory of Biophysics, Innovation Centre of Ministry of Education for Development and Diseases}, 
  \orgname{School of Medicine, South China University of Technology}, 
  \orgaddress{\city{Guangzhou}, \country{China}}}

\affil[2]{\orgdiv{Department of Radiology}, 
  \orgname{The First Affiliated Hospital of Sun Yat-sen University}, 
  \orgaddress{\city{Guangzhou}, \country{China}}}

\affil[3]{\orgdiv{Department of Radiology}, 
  \orgname{Guangzhou First People's Hospital, The Second Affiliated Hospital of South China University of Technology}, 
  \orgaddress{\city{Guangzhou}, \country{China}}}

\affil[4]{\orgdiv{Department of Neurosurgery and Radiology, College of Medicine}, 
  \orgname{Penn State College of Medicine}, 
  \orgaddress{\city{Hershey}, \state{Pennsylvania}, \country{USA}}}
  
  % 通讯作者（带 * 号）
\author*[1]{\fnm{Xuegang} \sur{Xin}}
\email{xinxg@scut.edu.cn}


\abstract{Noninvasive live-cell imaging in deep human tissues is crucial for exploring the cellular biological and pathogenic processes, but remains a significant unmet challenge. Diffusion magnetic resonance imaging (dMRI) promises to narrow this gap by noninvasively providing cellular-level microstructural information. Within a single crowded voxel containing millions of living cells, the intricate cellular-level microstructures create numerous microcompartments, each characterized by a specific diffusivity. However, conventional dMRI methods relying on voxel-averaged macroscopic parameters, merely reflect aggregate microstructural properties and fail to quantify this distribution of microcompartment-specific diffusivity within a voxel, thereby obscuring microstructural details. Here, we propose an intravoxel diffusivity probability distribution (IDPD) model to resolve a wealth of essential microstructural information via quantifying microcompartment-specific diffusivity distribution, thereby enabling direct cellular-level characterization. This exceptional capability is realized through a multi-tiered analytical workflow spanning targeted single-voxel or region of interest (ROI) analysis to global visualization using dynamic videos and statistic parametric maps. Ultimately, the IDPD model enables noninvasive cellular-level microstructure imaging, offering a promising avenue to evaluate living cell functions in vivo.}

\keywords{diffusion MRI, intravoxel diffusivity probability distribution, microstructure imaging}



\maketitle

\section{Introduction}\label{sec1}

Live-cell imaging is used by scientists to obtain a better understanding of  biological and pathogenic processes, usually performed with time-lapse microscopy on living cells cultured in vitro, but it cannot be used to study cellular functions in vivo\cite{bib1,bib2}. Intravital microscopy was invented for dynamic cellular-level imaging in vivo, however, it is typically invasive, which limits its ability to study intact cellular functions, and remains usually unsuitable for routine application within the human body\cite{bib3}. Noninvasively obtaining live-cell microscopic images to explore the biological and pathogenic processes in human tissues characterized by nontransparency in light is essentially meaningful in the field of life science in medicine, but it remains a significant unmet requirement in medical science.

Diffusion magnetic resonance imaging (dMRI) noninvasively provides unique microscopic insights into living cells within human tissues by detecting the displacements of diffusing water molecules\cite{bib4,bib5,bib6,bib7,bib8,bib9}. As noted by Professor Denis Le Bihan, the pioneer of dMRI, this technique can provide information on tissue at a microscopic level, while MR images remain at a macroscopic resolution. During a typical diffusion period spanning tens of milliseconds, dictated by the dMRI sequence, water diffusion displacements are distributed over micrometer-scale distances, which exactly match the dimensions of the cellular level\cite{bib5,bib6,bib7}. Water diffusion is restricted by cell membranes and hindered by intracellular and extracellular biomacromolecules\cite{bib5}. Additionally, biologically modulated water exchange across cell membranes contributes to water diffusion displacements\cite{bib6,bib7}. These cellular-level microstructures constitute the substrate underlying cellular functions. Consequently, cells undergoing dynamic cellular processes, such as activation, metabolism, proliferation, apoptosis, and necrosis, differentially alter their microenvironment, thereby exerting distinct impacts on water diffusion displacements\cite{bib10,bib11,bib12,bib13}.

Live-cell microscopic images acquired via light microscope typically feature a micrometer-scale resolution sufficient to distinguish individual cells, thereby enabling direct visualization of cellular morphology\cite{bib1,bib2,bib3}. However, macroscopic images obtained via dMRI are limited to a millimeter-scale resolution, meaning that each pixel in a dMRI image corresponds to a voxel containing millions of living cells\cite{bib4}.  Consequently, dMRI shifts its focus beyond morphological visualization to functional assessment by providing cellular-level microstructural information\cite{bib12,bib13}. Within such a crowded voxel, the intricate cellular-level microstructures create numerous microcompartments characterized by diverse diffusivity. These microcompartments do not necessarily correspond to specific anatomical spaces, but rather to functional water pools exhibiting similar diffusion behaviors. Thus, the heterogeneous cellular-level microstructures collectively result in a distribution of microcompartment-specific diffusivity within a single voxel.

Early dMRI introduced the apparent diffusion coefficient (ADC) to quantify water mobility based on the Gaussian diffusion assumption\cite{bib4}, where water diffusion displacements follow a Gaussian distribution. However, this simplified model aggregates intricate cellular-level microstructural information within a voxel into a single voxel-averaged macroscopic parameter. In vivo, however, the interaction between water molecules and microstructural barriers renders the displacement distribution intrinsically non-Gaussian. While diffusion tensor imaging (DTI) models diffusion anisotropy, particularly in organized tissues such as white matter (WM) and muscle, it remains constrained by the assumption of Gaussian diffusion along each direction\cite{bib14}. To bridge this gap, a variety of advanced models have been developed to characterize non-Gaussian diffusion behaviors\cite{bib15,bib16}, such as biexponential\cite{bib17,bib18}, kurtosis\cite{bib19}, stretched-exponential\cite{bib20}, multi-compartment\cite{bib21,bib22,bib23}, and statistical\cite{bib24,bib25,bib26,bib27} models. The biexponential model decomposes the dMRI signal into fast and slow diffusion components, quantified by respective diffusivities $(D_{\mathrm{f}} \text{ and } D_{\mathrm{s}})$\cite{bib18}. Nevertheless, the biophysical interpretation of these two components remains highly contentious\cite{bib28}. The kurtosis\cite{bib19} and stretched-exponential\cite{bib20} models characterize the deviation of water diffusion from Gaussian behavior through voxel-averaged parameters, namely the mean kurtosis (MK) and stretching parameter $\alpha$, respectively. However, these parameters offer a phenomenological description of microstructural heterogeneity, failing to resolve the underlying cellular-level microstructural information within a voxel. Multi-compartment models are primarily used to characterize brain microstructure by partitioning a voxel into a finite number of discrete anatomical spaces, e.g., intracellular, extracellular, and cerebrospinal fluid (CSF) compartments\cite{bib21,bib22,bib23}. These models enhance the biophysical interpretability of the estimated parameters by modeling the total dMRI signal as the weighted sum of signals from distinct compartments. However, the biophysical validity of these models remains strictly contingent upon a priori assumptions regarding the number, morphology, and orientation of the constituent compartments within a voxel. Unlike the preceding models predicated on tissue homogenization, the statistical models represent a transformative paradigm shift toward diffusivity distribution\cite{bib24,bib25,bib26,bib27}. By establishing a continuous probability density function (PDF) of diffusivity within a voxel, these models resolve the composite dMRI signal into a series of constituent diffusion components, each mapping to a specific microstructural environment. Ultimately, the emergence of this distributional paradigm enables a precise delineation of cellular-level microstructural information that conventional voxel-averaged parameters inevitably obscure. Despite its compelling theoretical underpinnings, the practical implementation of this paradigm remains challenging. Parametric approaches\cite{bib24,bib25}, while benefiting from computational simplicity by predefining functional forms (e.g., Gaussian, Gamma, or Beta distributions), often struggle to capture highly complex and dynamically evolving distributions. Conversely, non-parametric approaches offer the capacity to resolve arbitrary distributions\cite{bib26,bib27}. However, recovering a continuous distribution from finite, noisy measurements via the inverse Laplace transform (ILT) is a severely ill-posed inverse problem, wherein even minor measurement noise induces pronounced spurious peaks that obscure the true underlying distribution. 

Here, we first proposed an intravoxel diffusivity probability distribution (IDPD) model to resolve intricate cellular-level microstructural information within a single voxel. This model employs a novel iterative probability distribution approximation (IPDA) algorithm to achieve robust and high-fidelity quantification of IDPD. The proposed IDPD model was comprehensively validated via physical phantom imaging, healthy human brain imaging, and brain glioma imaging.

\section{Results}\label{sec2}
\subsection{Model validation via physical phantom imaging}

Validation using controlled aqueous phantoms demonstrated the capacity of the proposed IDPD model to distinguish intracellular from extracellular water diffusion in yeast cell suspensions in vitro (\textbf{Fig. 1}). The water phantom, exhibiting unrestricted free diffusion, yielded a sharp, unimodal IDPD profile (\textbf{Fig. 1a}). In contrast, cell suspensions containing intact yeast cells consistently produced broad, bimodal IDPD profiles characterized by the co-existence of low- and high-diffusivity components (\textbf{Fig. 1b-d}). Specifically, these bimodal profiles varied with cell density. With increasing cell density, from the upper (\textbf{Fig. 1b}) and lower (\textbf{Fig. 1c}) layers of a non-centrifuged cell suspension to a highly centrifuged cell suspension (\textbf{Fig. 1d}), the entire IDPD progressively shifted toward lower diffusivity regimes. Concurrently, this shift was accompanied by a monotonic increase in the probability of the low-diffusivity component and a corresponding decrease in that of the high-diffusivity component. To verify the microstructural origin of these two components, the yeast cells were subjected to ultrasonic disruption (\textbf{Fig. 1e}). Upon the loss of membrane integrity, the bimodal profile completely collapsed into a unimodal distribution positioned within the high-diffusivity regime. The disappearance of the low-diffusivity component definitively demonstrates that the low-diffusivity component is contingent upon membrane-enclosed compartmentalization, thereby resolving intracellular water diffusion. Conversely, the high-diffusivity component is attributed to the spaces between cells, reflecting extracellular water diffusion.

\subsection{ROI-based IDPD analysis of healthy human brain}

In vivo validation in the healthy human brain demonstrated that the IDPD model effectively differentiates cellular-level microstructural properties across distinct tissues, spanning from WM to cortical and subcortical gray matter (GM). The IDPD profiles at the population level revealed clear tissue-specific microstructural variations among eight representative ROIs (\textbf{Fig. 2a}). Nearly all tissues exhibited specific bimodal IDPD profiles. In highly anisotropic WM, the high-diffusivity component is strongly modulated by fiber orientation. Specifically, this component concentrated within higher diffusivity regimes along primary fiber orientations (e.g., Dir 1 for CC; Dir 2 for ALIC; Dir 3 for PLIC), but shifted toward reduced ranges along the orthogonal directions. Conversely, the high-diffusivity component in FWM containing crossing fibers exhibited diminished directional dependency. This variation results from the multi-directional orientation of crossing fibers within a single voxel. In contrast to these pronounced directional variations, the low-diffusivity component remains largely consistent across WM tissues, exhibiting only subtle fluctuations and remaining predominantly confined below 0.5 $\mu$m$^2$/ms. In GM tissues, the bimodal IDPD profiles were relatively isotropic across all directions. In these tissues, the low-diffusivity component remained primarily localized below 0.5 $\mu$m$^2$/ms, while the high-diffusivity component was predominantly concentrated between 0.5 and 1.5 $\mu$m$^2$/ms. 

In WM, the low-diffusivity component ($\le$ 0.5 $\mu$m$^2$/ms) is likely attributed to myelin water trapped within lamellar structures of myelin sheath. This diffusivity range aligns with reported myelin water ADC values (0.37 $\mu$m$^2$/ms parallel and 0.13 $\mu$m$^2$/ms perpendicular to fibers)\cite{bib29}, with radial diffusion being severely constrained by dense concentric lipid bilayers and the axolemma, while axial diffusion proceeds more freely along the longitudinal myelin sheath. Conversely, the high-diffusivity component (> 0.5 $\mu$m$^2$/ms) likely arises from coexisting intra- and extra-axonal water\cite{bib30,bib31,bib32}. Along the fiber orientation, axial diffusion within intra- and extra-axonal spaces proceeds with minimal restriction, experiencing weak hindrance primarily from longitudinal cytoskeletal elements (e.g., neurofilaments, microtubules) and the extracellular matrix. Conversely, radial diffusivity is substantially lower due to more pronounced microstructural constraints, specifically the intra-axonal confinement imposed by axonal membranes, and the elevated geometric tortuosity imposed on extra-axonal pathways by dense axonal packing.

In GM, the low- and high-diffusivity components likely associate with intracellular and extracellular water diffusion, respectively\cite{bib33,bib34}, while exhibiting negligible directional dependence. Notably, the low-diffusivity component is significantly higher in GM relative to WM, stemming from their disparate cellular microstructures. While water in WM is tightly restricted in myelinated axons, GM water resides within a larger, more isotropic intracellular space that permits higher diffusion freedom. Conversely, the high-diffusivity component is significantly lower in GM than that in highly anisotropic WM, exhibiting a trend observed in FWM. This reduction is primarily driven by the increased extracellular tortuosity.

\subsection{Voxel-based IDPD analysis of healthy human brain}

To evaluate whether the IDPD model can reliably capture cellular-level microstructural information at a single-voxel level, we examined three representative voxels extracted from each of the eight ROIs. \textbf{Fig. 2b} and \textbf{Extended Data Fig. 1a} respectively display the IDPD profiles for the first voxel and the remaining two voxels from each ROI, demonstrating the model's robustness in identifying tissue-specific properties and its sensitivity to subtle microstructural variations between individual voxels of the same tissue type. The measured and fitted dMRI signal decay of the corresponding voxels are displayed in \textbf{Fig. 2c} and \textbf{Extended Data Fig. 1b}. The high degree of agreement between experimental data and model fittings confirms that the estimated IDPDs accurately reflect the intrinsic microstructural properties of biological tissues, rather than representing fitting artifacts. Specifically, while the voxel-level IDPD profiles remained broadly consistent with the population-level trends, they exhibited noticeably narrower distribution widths for both IDPD components. Voxels within highly organized WM demonstrated pronounced orientation-dependent shifts in their IDPDs. In contrast, voxels within FWM exhibited less pronounced orientation dependence, due to the presence of crossing fibers. For cortical and subcortical GM, the voxel-wise IDPD profiles displayed substantial overlap across all directions, mirroring the isotropic nature of these tissues. Notably, even within the same tissue type, individual voxels exhibited discernible variations in the peak position, distribution width, and probability amplitude of both components.

\subsection{Comparison of IDPD and conventional voxel-averaged parameters}

To further evaluate the IDPD model's capacity to resolve the cellular-level microstructural information within a voxel, we performed a comparison between IDPD and conventional voxel-averaged parameters (\textbf{Fig. 3}), including ADC, FA, and mean diffusivity (MD) and MK, across twenty-four representative voxels ($n = 8$ per group, three groups). The IDPD model effectively deconstructs macroscopic parameter variations into distinct alterations of individual IDPD components. Specifically, rising FA values manifest as a marked inter-directional separation of IDPD components (\textbf{Fig. 3a}), thereby decomposing macroscopic anisotropy into direction-specific IDPD shifts. Elevated MK values are captured by a widening gap between IDPD components alongside enhanced inter-directional variability (\textbf{Fig. 3b}), revealing the biophysical substrates of non-Gaussian diffusion. Finally, increases in ADC or MD are primarily driven by a progressive shift of IDPD components toward higher diffusivity ranges (\textbf{Fig. 3c}), demonstrating that changes in voxel-averaged diffusivity fundamentally stem from the integrated alterations of local diffusivities.

These findings demonstrate that voxel-averaged parameters are merely coarse statistical descriptors of the underlying IDPD and thus suffer from degeneracy, in that, distinct microstructural properties can yield near-equivalent values. The IDPD model circumvents this degeneracy by resolving intricate microstructural details. For example, unlike FA, which frequently underestimates anisotropy and confounds crossing-fiber configurations with a loss of tissue integrity\cite{bib35}, the IDPD preserves microstructural sensitivity by resolving distinct fiber populations along individual directions. Furthermore, unlike MK, which aggregates all microstructural deviations, whether stemming from directional anisotropy or isotropic diffusivity variance\cite{bib19}, the IDPD resolves this ambiguity by deconstructing confounded sources into an explicit diffusivity distribution. Similarly, unlike ADC or MD, which are voxel-averaged diffusivities that inherently obscure the underlying microstructural details\cite{bib36}, the IDPD enables the explicit characterization of complex microstructural properties.

\subsection{IDPD visualization via dynamic video}

To characterize the spatial evolution of IDPDs across all voxels within each slice, we generated a dynamic video wherein each frame represents a probability map corresponding to a discrete diffusivity (\textbf{Extended Data Video 1}). This dynamic visualization effectively maps the underlying microstructural variations onto visualizable fluctuations in probabilities of specific diffusivities. We hypothesized that each discrete diffusivity corresponds to a specific class of microenvironment comprising identical or similar microstructural constituents. In each frame, the observed contrast reflects the varying fractional occupancies of a specific microenvironment across voxels. Crucially, adjacent voxels sharing the same microenvironmental signature coalesce into contiguous regions. As the diffusivity increments, the observational perspective toggles between different microenvironments, causing specific voxels to emerge or vanish. Visually, this manifests as a dynamic regional evolution, where the region boundaries appear to gradually extend outward or contract inward. Much like the movement of rain clouds in a weather forecast, where the presence of the cloud signifies the occurrence of rain. Through this dynamic visualization, each voxel is no longer a static data point but a fluid entity characterized by a unique evolutionary signature, facilitating the comprehensive resolution of heterogeneous microenvironments within each voxel. Of course, the precise microstructural origins of discrete diffusivity remain to be definitively established.

\subsection{IDPD visualization via static parametric maps}

To enable a straightforward comparison, we generated parametric maps for a total of sixteen parameters (\textbf{Fig. 4}), comprising three conventional voxel-averaged parameters (ADC, MD, and MK) and thirteen IDPD-derived parameters that capture the global and local statistical features of IDPD (detailed definitions are provided in Methods). The comparison of these parameters across eight representative ROIs is presented in \textbf{Extended Data fig. 2}.

The $\mu_D$ maps exhibited high consistency with conventional ADC and MD maps, which indirectly confirms the reliability of the IDPD model. However, unlike ADC and MD, which are constrained by the choice of $b$-values and cover only a limited diffusivity domain, $\mu_D$ bypasses these limitations by accounting for the entire diffusivity domain in a $b$-value-independent manner. Furthermore, MK merely quantifies the degree of deviation from Gaussian diffusion, the IDPD-derived parameters intrinsically dissect this overall deviation into more detailed microstructural information. Specifically, $\sigma_D$ measures the overall dispersion of IDPD. Elevated $\sigma_D$ values in WM tracts relative to GM tissues distinctly reflect greater microstructural heterogeneity inherent to axonal bundles. The $\mathrm{skew}_D$ quantifies the asymmetry of IDPD. Positive $\mathrm{skew}_D$ values are localized to highly aligned WM tracts (e.g., CC and PLIC), signifying an intra- or extra-axonal composition shifted toward high diffusivities. In contrast, other brain regions appear predominantly negative skewness. Concurrently, these WM tracts present the lowest $\mathrm{kurt}_D$ value, indicating a thin-tailed profile. Conversely, other regions exhibit heavier tails. Finally, $\sigma_P$ quantifies the sharpness of IDPD. Higher $\sigma_P$ values signify the sharper peaks, whereas lower $\sigma_P$ values reflect the smoother distribution. $\sigma_P$ serves as a complementary to $\sigma_D$. In multi-modal scenarios, $\sigma_D$ reflects the total distribution width, but is insensitive to the internal dispersion of individual IDPD components.

On $P_{\mathrm{low}}$ and $P_{\mathrm{mid}}$ maps, majority of brain tissues is relatively smooth, which is attributed to the high overlap of IDPD profiles within the low and medium diffusivity regimes. A notable exception occurs in highly aligned WM tracts (e.g., CC and PLIC), which manifest elevated $P_{\mathrm{low}}$ and correspondingly lower $P_{\mathrm{mid}}$ values specifically in directions perpendicular to the fiber orientation. Notably, for most other brain tissues, $P_{\mathrm{mid}}$ values are significantly higher than $P_{\mathrm{low}}$ values. The $\mathrm{MD}_{\mathrm{low}}$ maps exhibit a sharp contrast between GM and WM, with GM displaying significantly higher values compared to WM. This inter-tissue contrast notably diminishes on $\mathrm{MD}_{\mathrm{mid}}$ maps. On $P_{\mathrm{high}}$ and $\mathrm{MD}_{\mathrm{high}}$ maps, the CC and PLIC exhibit markedly higher values along the direction parallel to the fiber tracts. Beyond these well-organized WM structures, certain voxels surrounding the cerebrospinal fluid (CSF) also display moderate $P_{\mathrm{high}}$ and $\mathrm{MD}_{\mathrm{high}}$ values, revealing prominent partial volume effects. The $P_{\mathrm{free}}$ and $\mathrm{MD}_{\mathrm{free}}$ maps exclusively delineate CSF-filled regions, serving as a highly specific identifier for free water. Due to the flow effects of CSF, the $\mathrm{MD}_{\mathrm{free}}$ values exceed the diffusivity of free water at 37 $^\circ$C.

\subsection{Brain glioma imaging reveals clinical utility of IDPD model}

To evaluate the clinical utility of the IDPD model, we extended in vivo validation to brain gliomas imaging. We summarized the mean IDPDs averaged across all voxels within four pathologically distinct ROIs in brain of a high-grade glioma patient (\textbf{Fig. 5a}), while the IDPDs of three representative voxels extracted from each of the four ROIs (\textbf{Fig. 5b} and \textbf{Extended Data Fig. 3a}). The measured and fitted dMRI signal decays of the corresponding voxels are displayed in \textbf{Fig. 5c} and \textbf{Extended Data Fig. 3b}. Within the enhanced solid tumor region (ROI 1), the IDPD profiles are near-identical across all directions. This isotropic distribution signifies that aggressive tumor proliferation has effectively destroyed the anisotropic WM architecture, replacing it with a disorganized, hypercellular microenvironment. Conversely, within the tumor-infiltrated association fibers (ROI 2), the IDPD manifests a significant directional dependency. The IDPD along Dir 2 demonstrates a pronounced shift toward higher diffusivity compared to ROI 1, whereas the IDPDs along Dir 1 and Dir 3 exhibits a slight bias toward lower diffusivity. This finding suggests that despite tumor infiltration, residual axonal barriers remain intact to modulate water diffusion. Within the tumor-infiltrated optic radiation (ROI 3), the IDPDs manifest a distinct signature where the low-diffusivity component exhibits a shift toward even lower diffusivity, whereas the high-diffusivity component shows no pronounced elevation across directions. This behavior could be attributed to the mass effect of the tumor, which induces dense axonal packing and reduced volume of extracellular space. The non-enhancing necrotic core (ROI 4) is dominated by the high-diffusivity components, confirming liquefactive necrosis where the organized tissue architecture has disintegrated into fluid-filled cavities. Within this region, water diffusion is hindered only by macromolecular remnants and sparse cellular debris.

The dynamic spatial evolution of IDPDs across all voxels is captured in \textbf{Extended Data Video 2}, with corresponding parametric maps presented in \textbf{Fig. 6}. The distinct pathological regions, such as tumor infiltration, peritumoral edema, necrosis, or cystic transformation, exhibit significantly different evolutionary patterns, allowing their respective boundaries to be observed intuitively. The parametric maps represent a synthesized result. The diversity of parameters undoubtedly facilitates more precise disease assessment, as demonstrated by the significant differences across ROIs (\textbf{Extended Data fig. 4}). For instance, from ROI 1 to ROI 3, higher fiber tract integrity correlates with an elevated $P_{\mathrm{low}}$ and a reduced $\mathrm{MD}_{\mathrm{low}}$, suggesting that these two parameters could serve as sensitive indicators for WM tracts infiltrated by tumor cells. Meanwhile, the highest $P_{\mathrm{high}}$ and $\mathrm{MD}_{\mathrm{high}}$ values are observed in ROI 4, whereas noticeably lower values are found in ROI 1 and ROI 2, highlighting their potential as imaging surrogates for characterizing edema or necrosis. Finally, from ROI 1 to ROI 3, $P_{\mathrm{mid}}$ consistently exceeds 0.5, a value substantially higher than that in ROI 4, indicating that $P_{\mathrm{mid}}$ and $\mathrm{MD}_{\mathrm{mid}}$ hold promise as sensitive correlates for evaluating cellular density.

\section{Discussion}\label{sec3}

In this work, the proposed IDPD model, supported by the IPDA algorithm, exhibits a remarkable capability to resolve a wealth of essential cellular-level microstructural information within a single voxel. By substituting voxel-averaged parameters, such as ADC, MD, and MK, with a diffusivity distribution, this model successfully overcomes the traditional dMRI limitation of reflecting only aggregate microstructural properties. By bridging the gap between macroscopic imaging and cellular-level microstructural information, the IDPD model provides a promising pathway for cellular-level microstructure imaging, offering deeper insights into cell dynamics and function in vivo. Crucially, the IDPD model holds high translational potential for routine clinical practice. 

The practical utility of the IDPD model is realized through a multi-tiered analytical workflow designed to meet the distinct demands of both clinical use and scientific research. Initially, the global visualization via dynamic video facilitates rapid screening and identification of suspicious lesions. Subsequently, the global visualization via static parametric maps permits precise localization of lesions combined with routine structural MRI images. Furthermore, the ROI-based assessment allows for quantitative analysis of both IDPD-derived parameters and entire IDPD profiles within selected lesion regions. Ultimately, the voxel-wise analysis achieves an exhaustive evaluation of subtle pathological alterations at single-voxel level.

A single dMRI voxel in the human brain integrates dMRI signals from distinct water microcompartments governed by intricate microstructures at the cellular and subcellular levels. Even under normal physiological homeostasis, this biophysical compartmentalization is inherently intertwined, varying significantly across distinct tissues and among discrete voxels within the same tissue. Pathological insults, such as vasogenic edema, tumor infiltration, or liquefactive necrosis, further obliterate microstructural boundaries, exponentially amplifying the microstructural complexity. Conventional dMRI methods collapse this complexity into voxel-averaged parameters, thereby introducing biophysical degeneracy. For instance, the reduction in FA can ambiguously arise from demyelination, axonal degradation, increased fiber crossing, or isolated vasogenic edema, which frequently confounds the assessment of tissue integrity\cite{bib38}. Similarly, although ADC is employed as imaging surrogate for tumor cellularity, empirical findings across studies remain inconsistent and contradictory\cite{bib39}. The IDPD model ensures that subtle intravoxel pathological alterations remain highly discernible by resolving the rich microstructural details within a single voxel, as shown in \textbf{Fig. 5a, b} and \textbf{Fig. 6}. The diffusivity distribution below 0.5 $\mu$m$^2$/ms holds potential as surrogate for demyelination, neurodegeneration, or tumor-infiltrated tracts, bypassing the confounding effects inherent to FA. The middle diffusivity regime (from 0.5 to 1.5 $\mu$m$^2$/ms) typically contributes the primary dMRI signal and may provide a sensitive correlate for cellular density. The higher diffusivities (from 1.5 to 3.0 $\mu$m$^2$/ms) capture the fast-diffusing extracellular water under conditions of edema or necrosis, in which altered macromolecular networks exert only weak hindrance on water. The diffusivities beyond 3.0 $\mu$m$^2$/ms purely isolate unhindered, isotropic free water, providing a clean mapping of CSF. Although the precise correlations between specific diffusivity and complex microcompartments warrants ongoing validation through extensive, rigorous efforts across the dMRI community, the IDPD model overcomes the inherent biophysical degeneracy of voxel-averaged parameters. By doing so, the IDPD model pioneers a novel diagnostic modality that establishes a single voxel as a standalone vehicle for comprehensive assessment of cellular-level biological and pathogenic processes.

Although the IDPD model shares the foundational objective of quantifying intravoxel diffusivity distributions with established statistical models, it is a distinct framework designed to circumvent their inherent limitations. To date, the clinical translation of conventional statistical models has been severely hampered by restrictive model assumptions, algorithmic limitations, and constrained representation modes\cite{bib24,bib25,bib26,bib27}. Existing parametric statistical models, such as those assuming truncated Gaussian or gamma distributions, inevitably rely on voxel-averaged parameters (e.g., mean and standard deviation for Gaussian distribution, or shape and scale parameters for gamma distribution)\cite{bib24,bib25}. Consequently, these parameters have offered no clear diagnostic superiority over parameters derived from monoexponential or non-monoexponential models\cite{bib40,bib41}. In contrast, through a multi-tiered analytical workflow, the IDPD model provides critical microstructural insights that remain inaccessible via voxel-averaged parameters. Furthermore, while non-parametric statistical models facilitate the quantification of arbitrary distributions beyond simplified parametric forms, their clinical adoption is severely bottlenecked by algorithmic limitations. The ILT underpinning these models is highly sensitive to measurement noise, frequently yielding non-unique or biophysically implausible solutions\cite{bib26,bib27}. Conversely, the IPDA algorithm achieves robust quantification characterized by its remarkable resistance to noise. The high consistency between the fitted and measured data confirms superior goodness-of-fit of IPDA algorithm, while ensuring that the resolved distributions remain universally stable and biophysically interpretable across diverse tissue types.

Several limitations of the IDPD model warrant consideration. First, the explicit biophysical link between local diffusivity and microstructures remains to be fully elucidated, currently constrained by the lack of established third-party technologies capable of measuring water diffusion at comparable spatial scales\cite{bib42}. Future validation via advanced optical microscopy or histopathology is required to establish a definitive microstructural ground-truth. Second, this study utilized a fixed long diffusion time via standard pulsed gradient spin-echo (PGSE) sequence, which overlooks the time-dependency of diffusivity and limits the model's sensitivity to small microstructures. Future work should incorporate oscillating gradient spin-echo (OGSE)\cite{bib43} or custom-designed sequences to achieve shorter diffusion times and broader temporal coverage. Third, although the current 46 b-values acquisition scheme is essential for methodological establishment, it prolongs scan time for routine clinical use. Identifying a minimal subset of b-values is crucial to reduce scan time while maintaining robust quantification. Finally, the IDPD model is inherently generalizable across diverse pathological conditions, warranting future validation through large-scale, multicenter cohort studies.

\section{Methods}\label{sec4}
\subsection{Modeling: IDPD model}

For Gaussian diffusion in homogeneous media, the diffusivity is considered uniform within a voxel, yielding a mono-exponential dMRI signal decay \(S/S_0\):
\begin{equation}
\frac{S}{S_0} = \exp(-b \cdot \mathrm{ADC}) \label{eq:1}
\end{equation}
where \(b\)-value is the diffusion weighted factor, and \(\mathrm{ADC}\) represents the voxel-averaged diffusivity. \(S_0\) and \(S\) represent the dMRI signal acquired with \(b=0\) and \(b \neq 0~\mathrm{ms}/\mu\mathrm{m}^2\), respectively.

In contrast, in biological tissues, water molecules exhibit non-Gaussian diffusion behavior. A voxel representing a pixel in a dMRI image encompasses millions of cells, whose complex microstructures create numerous microscopic compartments characterized by diverse water mobility. It should be noted that these microcompartments discussed herein do not necessarily correspond to specific anatomical spaces, but rather to functional water pools exhibiting similar diffusion behaviors or identical diffusivity. Consequently, the diverse cellular-level microstructures (e.g., cell membranes, organelles, nuclei, and the extracellular matrix) impose distinct physical barriers to water molecules, yielding a heterogeneous distribution of microcompartment-specific diffusivity within a voxel.

Given the continuous variations in cellular geometry and density, this intravoxel diffusivity distribution is statistically smooth. Mathematically, it is characterized by a continuous probability density function (PDF) of diffusivity, \(f(D)\), where \(D\) denotes the local diffusivity. Assuming Gaussian diffusion within each microcompartment, which yields a local mono-exponential signal decay, the total signal decay, \(S(b_m)/S_0\), for a voxel at a given nonzero \(b\)-value, \(b_m\), can be expressed as the integral of exponential decay kernels, with \(f(D)\) as the weighting function:
\begin{equation}
\frac{S(b_m)}{S_0} = \int_{D_{\min}}^{D_{\max}} f(D) \cdot \exp\left(-b_m \cdot D\right) \, dD, \quad m \in \{1,2,\dots, M_b\} \label{eq:2}
\end{equation}
where \(m\) and \(M_b\) are the index and number of nonzero \(b\)-values, respectively. \(D_{\min}\) and \(D_{\max}\) represent the minimal and maximal diffusivity within a voxel.

The universal approximation capability of Gaussian mixture model (GMM) suggests that any continuous PDF can be well-approximated by a linear combination of Gaussian basis functions\cite{bib44}. Motivated by this principle in statistics, \(f(D)\) is modeled as a mixture of \(J\) Gaussian components:
\begin{equation}
f(D) = \sum_{j=1}^{J} \pi_j \cdot \mathcal{N}\!\left(D \mid \mu_j, \sigma_j^2\right) \label{eq:3}
\end{equation}
where \(\pi_j\) represents the mixture weight of the \(j\)-th Gaussian component, and \(\mathcal{N}\!\left(D \mid \mu_j, \sigma_j^2\right)\) denotes the \(j\)-th normal distribution with mean \(\mu_j\) and standard deviation \(\sigma_j\). GMM requires these weights to be positive (\(\pi_j > 0\)) and normalized such that \(\sum_{j=1}^{J} \pi_j = 1\). However, solving for the continuous \(f(D)\) is challenged by the non-negativity of diffusivity (\(D > 0\)). To address this challenge, the continuous PDF estimation is reformulated as a discrete probability distribution estimation. The continuous diffusivity domain between \(D_{\min}\) and \(D_{\max}\) is discretized into \(N_D\) equally spaced diffusivity intervals. Each interval, indexed by \(n \in \{1,2,\dots, N_D\}\), corresponds to an interval-averaged diffusivity, \(D_n\). Thus, equation \eqref{eq:2} is discretized as the following form:
\begin{equation}
\frac{S(b_m)}{S_0} = \sum_{n=1}^{N_D} p(D_n) \cdot \exp\!\left(-b_m \cdot D_n\right) \label{eq:4}
\end{equation}
where \(\{p(D_n)\}_{n=1}^{N_D}\) (denoted as \(\mathcal{P}_D\)) signifies the probability distribution across all \(D_n\) values, and satisfies \(\sum_{n=1}^{N_D} p(D_n) = 1\), and \(p(D_n) > 0\). Discrete probability \(p(D_n)\) equals the integral of \(f(D)\) over the \(n\)-th interval:
\begin{equation}
p(D_n) = \int_{D_n - \frac{\Delta D}{2}}^{D_n + \frac{\Delta D}{2}} f(D) \, dD \label{eq:5}
\end{equation}
where \(\Delta D\) represents the interval width. While a smaller \(\Delta D\) facilitates a finer discretization of \(f(D)\), it thereby enables \(\mathcal{P}_D\) to characterize the inherent probability distribution, it comes at the expense of requiring denser \(b\)-value sampling and increased computational overhead. In this study, \(\Delta D\) was set to \(0.025~\mu\mathrm{m}^2/\mathrm{ms}\), with discrete \(D_n\) values defined at the interval midpoints (e.g., 0.0125, 0.0375, 0.0625, \(\ldots~\mu\mathrm{m}^2/\mathrm{ms}\)). Accordingly, the ADC can be defined as the weighted sum of local diffusivities across all microcompartments, expressed as:
\begin{equation}
\mathrm{ADC} = \int_{D_{\min}}^{D_{\max}} f(D) \cdot D \, dD \approx \sum_{n=1}^{N_D} p(D_n) \cdot D_n \label{eq:6}
\end{equation}

The IDPD model aims to estimate the inherent probability distribution \(\mathcal{P}_D\) within each voxel, thereby resolving the intricate microstructural properties.

\subsection{Fitting: IPDA algorithm}

To address the challenge of robustly estimating \(\mathcal{P}_D\), we introduced an IPDA algorithm, with its implementation details summarized in \textbf{Table 1}. This algorithm adaptively updates the prior probability distribution, \(\{p_{\mathrm{prior}}(D_n)\}_{n=1}^{N_D}\) (denoted as \(\mathcal{P}_{\mathrm{prior}}\)), in each iteration. The updated \(\mathcal{P}_{\mathrm{prior}}\) then guides the data-driven approximation of the target probability distribution \(\mathcal{P}_D\) via a regularized non-negative least-squares (NNLS) estimation.

\subsection{IPDA algorithm: initial optimization}

The construction of the initial prior probability distribution, \(\mathcal{P}_{\text{prior}}^{(0)}\), leverages the \(b\)-value dependence of ADC. In biological tissue, increasing the \(b\)-value amplifies diffusion weighting, preferentially suppressing the dMRI signal from high-diffusivity microcompartments, and thereby driving a progressive increase in the relative signal contribution from low-diffusivity microcompartments. Consequently, ADC exhibits a monotonic decrease with increasing \(b\)-value, a trend reflecting the shift in the underlying probability distribution. To quantify the trend, the \(\mathrm{ADC}_m\) value for each non-zero \(b\)-value \((b_m)\) is calculated, via equation (1):

\begin{equation}
\mathrm{ADC}_m = \frac{-\ln\!\left(\frac{S(b_m)}{S_0}\right)}{b_m} \label{eq:7}
\end{equation}

According to equation (6), \(\mathrm{ADC}_m\) can be further expressed as:

\begin{equation}
\mathrm{ADC}_m = \int_{D_{\min}}^{D_{\max}} f_m(D) \cdot D \, dD \approx \sum_{n=1}^{N_D} p_m(D_n) \cdot D_n \label{eq:8}
\end{equation}
where \(f_m(D)\) and \(\{p_m(D_n)\}_{n=1}^{N_D}\) (denoted as \(\mathcal{P}_m\)) denote the continuous PDF and the discrete probability distribution of diffusivity at a given \(b_m\), respectively. Crucially, \(f_m(D)\) and \(\mathcal{P}_m\) serve as the \(b\)-value-dependent manifestation of the inherent \(f(D)\) and \(\mathcal{P}_D\). As \(b_m\) increases, the reduction in \(\mathrm{ADC}_m\) signifies a ``left-shift'' of \(\mathcal{P}_m\) toward lower diffusivity. Specifically, starting with the highest-diffusivity interval, the \(p_m(D_n)\) progressively decreases to zero, while maintaining \(\sum_{n=1}^{N_D} p_m(D_n) = 1\).

We assume that each \(f_m(D)\) follows a normal distribution \(\mathcal{N}_m\!\left(D \mid \mathrm{ADC}_m, \sigma_m^2\right)\) with mean \(\mathrm{ADC}_m\) and standard deviation \(\sigma_m\). Following the GMM, the initial prior PDF \(f^{(0)}(D)\) can be modeled as a mixture of \(\mathcal{N}_m\!\left(D \mid \mathrm{ADC}_m, \sigma_m^2\right)\) across all non-zero \(b\)-values:

\begin{equation}
f^{(0)}(D) = \sum_{m=1}^{M_b} \pi_m \cdot \mathcal{N}_m\!\left(D \mid \mathrm{ADC}_m, \sigma_m^2\right) \label{eq:9}
\end{equation}
where \(\pi_m\) represents the mixture weight of the \(m\)-th Gaussian component. In the absence of prior information regarding \(\pi_m\), for simplicity, we initialize all Gaussian components with an identical weight \(\pi_m = 1/M_b\). These initial weights are employed only for the construction of \(f^{(0)}(D)\).

Grounded in the assumption that measurement uncertainty is equivalent across all \(b\)-values, all Gaussian components share an identical standard deviation. More importantly, imposing a constant \(\sigma_m\) constraint serves as an essential regularization mechanism to address the ill-posed nature of probability distribution estimation and prevent numerical singularities. If individual standard deviations are allowed to vary independently, the IDPD model would tend to overfit the localized measurement noise, causing corresponding \(\sigma_m\) to vanish (\(\sigma_m \to 0\)) and thereby resulting in spurious, noise-driven sharp peaks. By assigning an appropriate, uniform \(\sigma_m\) that forces each Gaussian component to maintain a finite width, ultimately, the IPDA algorithm effectively eliminates such singularities and renders the ill-posed inverse problem well-posed, thereby ensuring the computational robustness. \(\sigma_m\) is determined adaptively via Silverman's Rule of Thumb\cite{bib45}. This rule is adopted because its closed-form solution significantly reduces the computational burden of voxel-wise fitting, while its optimization based on asymptotic mean integrated squared error exerts an inherent smoothing effect that effectively suppresses spurious noise-induced artifacts. Specifically, \(\sigma_m\) is formulated as follows:

\begin{equation}
\sigma_m = 1.06 \cdot \min\!\left(\sigma_{\mathrm{ADC}}, \frac{\operatorname{IQR}(\mathrm{ADC}_m)}{1.34}\right) \cdot M_b^{-1/5} \label{eq:10}
\end{equation}
where \(\min(\cdot)\) selects the smaller value of its arguments, and \(\operatorname{IQR}(\mathrm{ADC}_m)\) denotes the interquartile range of \(\mathrm{ADC}_m\) values. The \(\sigma_{\mathrm{ADC}}\) represents the weighted standard deviation of \(\mathrm{ADC}_m\) values:

\begin{equation}
\sigma_{\mathrm{ADC}} = \sqrt{\sum_{m=1}^{M_b} \pi_m \cdot \left(\mathrm{ADC}_m - \sum_{m=1}^{M_b} \pi_m \cdot \mathrm{ADC}_m\right)^2} \label{eq:11}
\end{equation}

In resolving the ill-posed problem, it is of paramount importance to determine the limits of \(D_n\) values within each voxel. Fundamentally, these limits inherently reflect the detectable diffusivity domain under the employed \(b\)-value scheme. Accordingly, strictly restricting the solution space of the ill-posed problem to a finite, data-derived diffusivity domain guarantees the biophysical interpretability of estimated \(\mathcal{P}_D\). This restriction further precludes implausible solutions induced by measurement noise. The minimum and maximum \(\mathrm{ADC}_m\) values, denoted as \(\mathrm{ADC}_{\min}\) and \(\mathrm{ADC}_{\max}\), provide a practical basis for determining the lower and upper limits of \(D_n\) values, denoted as \(D_{\min}^{(0)}\) and \(D_{\max}^{(0)}\). Given that each \(\mathrm{ADC}_m\) represents the mean of a normal distribution, to rigorously minimize the truncation errors, the lower and upper limits of \(D_n\) values are established to encompass \(>99.999\%\) of the underlying diffusivity domain (which corresponds to a five-sigma span), as follows:

\begin{equation}
\begin{split}
D_{\min}^{(0)} &= \max\!\left(0, \; \mathrm{ADC}_{\min} - 5\sigma_m\right) \\
D_{\max}^{(0)} &= \mathrm{ADC}_{\max} + 5\sigma_m
\end{split}
\label{eq:12}
\end{equation}
where \(\max(0,\dots)\) selects the larger value of its arguments to prevent negative diffusivity. To facilitate computation, these limits are rounded to the discrete \(D_n\) values, which then dictate the initial number of intervals, \(N_D^{(0)}\).

Building upon the established \(f^{(0)}(D)\), the \(\mathcal{P}_{\text{prior}}^{(0)}\) is constructed by integrating \(f^{(0)}(D)\) across \(N_D^{(0)}\) diffusivity intervals centered at \(D_n \in [D_{\min}^{(0)}, D_{\max}^{(0)}]\). The prior probability \(p_{\text{prior}}^{(0)}(D_n)\) for the \(n\)-th interval is calculated as follows:

\begin{equation}
p_{\text{prior}}^{(0)}(D_n) = \int_{D_n - \frac{\Delta D}{2}}^{D_n + \frac{\Delta D}{2}} f^{(0)}(D) \, dD \label{eq:13}
\end{equation}

Theoretically, a sufficiently broad and dense \(b\)-value sampling could exhaustively capture the dMRI signal contributions from all intrinsic diffusivities within a voxel. Under such idealized conditions, \(\mathcal{P}_{\text{prior}}^{(0)}\) would asymptotically converge to the inherent \(\mathcal{P}_D\). However, in clinical practice, \(b\)-value coverage and sampling density are strictly constrained by permissible scan durations. Furthermore, the weights \(\pi_m\) of individual Gaussian components are intrinsically variable. Consequently, \(\mathcal{P}_{\text{prior}}^{(0)}\) serves merely as a theoretical approximation. More importantly, relying solely on this prior is inadequate to accurately characterize the measured dMRI signal decay.

To rigorously fit the empirical data, an NNLS estimation is introduced. In this process, \(\mathcal{P}_{\text{prior}}^{(0)}\) is utilized to explicitly constrain the solution space, functioning as a physics-informed regularization. Concurrently, to enforce the non-negativity of probabilities, each estimated probability \(p(D_n)\) is subject to the following bounds during the subsequent regularized NNLS estimation:

\begin{equation}
0 \leq p(D_n) \leq \min\!\left(1, \; p_{\text{prior}}^{(0)}(D_n) + p_c^{(0)}(D_n)\right) \label{eq:14}
\end{equation}

Within this constraint, a uniform compensation term, \(\{p_c^{(0)}(D_n)\}_{n=1}^{N_D^{(0)}}\) (denoted as \(\mathcal{P}_c^{(0)}\)), is introduced. By uniformly shifting \(\mathcal{P}_{\text{prior}}^{(0)}\) upward, this term preserves the essential distributional profile while affording each estimated \(p(D_n)\) sufficient flexibility to deviate from the prior. Because the prior varies intrinsically across individual voxels, applying a globally fixed compensation would yield inconsistent constraint strengths. To adaptively address this, \(p_c^{(0)}(D_n)\) is defined as the voxel-specific mean of the prior probabilities (i.e., \(p_c^{(0)}(D_n) = \frac{1}{N_D^{(0)}} \sum_{n=1}^{N_D^{(0)}} p_{\text{prior}}^{(0)}(D_n)\)). Concurrently, the \(\min(1,\dots)\) operator rigorously ensures that the theoretical upper bound of any individual probability does not exceed unity. With the feasible solution space strictly bounded by equation (14), we implement the following regularized NNLS estimation:

\begin{equation}
\mathcal{P}_D^{(0)} = \underset{p(D_n)}{\operatorname{argmin}} \; \sum_{m=1}^{M_b} \left[ \sum_{n=1}^{N_D^{(0)}} p(D_n) \cdot \exp\!\left(-b_m \cdot D_n\right) - \frac{S(b_m)}{S_0} \right]^2 \label{eq:15}
\end{equation}

This objective function in equation (15) quantifies the squared residual error between the fitted and measured dMRI signals. Minimizing this function yields the initial probability distribution, \(\{p^{(0)}(D_n)\}_{n=1}^{N_D^{(0)}}\) (denoted as \(\mathcal{P}_D^{(0)}\)). Unlike \(\mathcal{P}_{\text{prior}}^{(0)}\), which is a theoretical construct, \(\mathcal{P}_D^{(0)}\) represents the empirically derived probability distribution that faithfully reflects the dMRI signal decay. In essence, this regularized NNLS estimation refines the purely theoretical approximation of the inherent \(\mathcal{P}_D\) in a data-driven manner.

Accurate estimation of the inherent \(\mathcal{P}_D\) relies on a robust prior, making the fidelity of the mixture weights paramount. However, because the construction of \(\mathcal{P}_{\text{prior}}^{(0)}\) does not account for variations in the mixture weights of individual Gaussian components, the initially estimated \(\mathcal{P}_D^{(0)}\) inevitably deviates from the inherent \(\mathcal{P}_D\). This deviation is particularly pronounced in complex distributions characterized by heavy-tailed, asymmetric, or multimodal profiles. To overcome this limitation, the key to the subsequent iterative optimization lies in the adaptive optimization of mixture weights.

\subsection{IPDA algorithm: iterative optimization}

Specifically, the probability distribution derived from the preceding regularized NNLS estimation is used to refine the prior, which subsequently guides the current estimation, thereby driving a self-reinforcing optimization cycle. The iteration starts with \(\mathcal{P}_D^{(0)}\). Given that \(\mathcal{P}_D^{(0)}\) is defined over discrete \(D_n\) values, the original \(\mathrm{ADC}_m\) are substituted with \(D_n\) to formulate the Gaussian mixture.

For the \(k\)-th iteration, the continuous PDF, \(f^{(k)}(D)\), is formulated as a weighted mixture of new Gaussian components \(\mathcal{N}_n\!\left(D \mid D_n, \sigma_n^{(k)^2}\right)\) with mean \(D_n\) and standard deviation \(\sigma_n^{(k)}\):

\begin{equation}
f^{(k)}(D) = \sum_{n=1}^{N_D^{(k-1)}} \pi_n^{(k)} \cdot \mathcal{N}_n\!\left(D \mid D_n, \sigma_n^{(k)^2}\right) \label{eq:iter16}
\end{equation}
where \(\pi_n^{(k)}\) represents the mixture weight of the \(n\)-th Gaussian component in the \(k\)-th iteration. The normalization of the mixture distribution \(f^{(k)}(D)\) requires its integral over the diffusivity domain to equal unity:

\begin{equation}
\int_{D_{\min}^{(k)}}^{D_{\max}^{(k)}} f^{(k)}(D) \, dD = \int_{D_{\min}^{(k)}}^{D_{\max}^{(k)}} \sum_{n=1}^{N_D^{(k-1)}} \pi_n^{(k)} \cdot \mathcal{N}_n\!\left(D \mid D_n, \sigma_n^{(k)^2}\right) dD = \sum_{n=1}^{N_D^{(k-1)}} \pi_n^{(k)} = 1 \label{eq:iter17}
\end{equation}

Given that the previously derived \(\mathcal{P}_D^{(k-1)}\) satisfies \(\sum_{n=1}^{N_D^{(k-1)}} p^{(k-1)}(D_n) = 1\), a deterministic one-to-one relationship is established: \(\pi_n^{(k)} = p^{(k-1)}(D_n)\). Consequently, the \(n\)-th weighted Gaussian component contributes a total probability equal to its weight \(\pi_n^{(k)}\), effectively scaling its contribution to the mixture distribution while maintaining its original mean.

The \(\sigma_n^{(k)}\) is still updated via equation (10) by replacing the parameters \(\sigma_{\mathrm{ADC}}\), \(\operatorname{IQR}(\mathrm{ADC}_m)\), and \(M_b\) with \(\sigma_D\), \(\operatorname{IQR}(D_n)\), and \(N_D^{(k-1)}\), respectively. Here, \(\sigma_D\) denotes the weighted standard deviation of \(D_n\) values, which is updated via equation (11) by analogously substituting \(D_n\), \(\pi_n^{(k)}\), and \(N_D^{(k-1)}\) for \(\mathrm{ADC}_m\), \(\pi_m\), and \(M_b\). The \(\operatorname{IQR}(D_n)\) represents the weighted interquartile range of \(D_n\) values. Following the update of \(\sigma_n^{(k)}\), each Gaussian component is determined. These components, along with their updated weights \(\pi_n^{(k)}\), are then integrated according to equation (16) to yield the refined \(f^{(k)}(D)\).

Although the limits of \(D_n\) values have been initially determined via equation (12), this calculation is predicated on the initial \(\pi_m\) and \(\sigma_m\). The updated \(\pi_n^{(k)}\) and \(\sigma_n^{(k)}\) necessitate a further refinement of these limits to accurately reflect the underlying diffusivity domain. According to equation (17), the lower limit corresponds to the minimum diffusivity of the first Gaussian component centered at \(D_{\min}^{(k-1)}\) within the mixture distribution, whereas the upper limit is defined by the maximum diffusivity within the last Gaussian component centered at \(D_{\max}^{(k-1)}\). Let \(\pi_{\min}^{(k)}\) and \(\pi_{\max}^{(k)}\) respectively represent the weights of the two Gaussian components. The updated limits, denoted as \(D_{\min}^{(k)}\) and \(D_{\max}^{(k)}\), are then determined as follows:

\begin{equation}
\begin{split}
D_{\min}^{(k)} &= \max\!\left(0, \; D_{\min}^{(k-1)} - \sigma_n \Phi^{-1}\!\left(\frac{1 + \pi_{\min}}{2}\right)\right), \\
D_{\max}^{(k)} &= D_{\max}^{(k-1)} + \sigma_n \Phi^{-1}\!\left(\frac{1 + \pi_{\max}}{2}\right).
\end{split}
\label{eq:18}
\end{equation}
where \(\Phi^{-1}(\cdot)\) denotes the inverse cumulative distribution function of the standard normal distribution. Like the initial limits, these updated limits are rounded to the discrete \(D_n\) values. Intuitively, the extent of expansion scales with the component weights, with the limits asymptotically converging to their current values as weights vanish. Notably, equation (18) cannot supersede equation (12) for the initial determination of these limits, as the initial weights are manually assigned. Similarly, equation (12) is also unsuitable for iterative refinement as it would result in excessive expansion. Accordingly, the number of intervals \(N_D^{(k)}\) is updated.

The updated prior, \(\mathcal{P}_{\text{prior}}^{(k)}\), is then obtained by integrating \(f^{(k)}(D)\) across \(N_D^{(k)}\) diffusivity intervals, providing a basis to refine the probability bounds via equation (14). Finally, a regularized NNLS estimation is performed according to equation (15) to update the empirically derived probability distribution, \(\{p^{(k)}(D_n)\}_{n=1}^{N_D^{(k)}}\) (denoted as \(\mathcal{P}_D^{(k)}\)).

As the iterations progress, the weights \(\pi_n^{(k)}\) are progressively refined. Consequently, the estimated \(\mathcal{P}_D^{(k)}\) converges toward the inherent \(\mathcal{P}_D\), eventually reaching a stable state. The iterative process terminates either when the sum of absolute differences between \(\mathcal{P}_D^{(k-1)}\) and \(\mathcal{P}_D^{(k)}\) falls below \(1 \times 10^{-3}\), or upon reaching a maximum of ten iterations to prevent infinite looping. The final estimated probability distribution is then established as the inherent \(\mathcal{P}_D\).

\subsection{Participants and phantoms}

This study was approved by the Institutional Review Board of The First Affiliated Hospital, Sun Yat-Sen University, and written informed consent was obtained from all participants. The study population included seven healthy volunteers and seven histologically confirmed glioma patients (three WHO grade II astrocytoma, one WHO grade IV astrocytoma, and three WHO grade IV glioblastomas).

To validate the IDPD model, four aqueous phantoms were prepared in $50~\mathrm{mL}$ centrifuge tubes to simulate distinct diffusion environments. A water phantom was prepared using deionized water to serve as a free diffusion control. The remaining three phantoms were prepared using a 2:1 (w/w) yeast suspension in deionized water, processed into three states: non-centrifuged, centrifuged, and a centrifuged suspension after cell disruption (VCX 500, Sonics \& Materials, Newtown, USA) to simulate varying degrees of cell density and integrity.

\subsection{Magnetic resonance imaging}

Imaging was performed on a 3T scanner (Prisma, Siemens Healthcare, Erlangen, Germany) equipped with an 80~mT/m gradient system and a 64-channel head/neck coil. The dMRI signals for IDPD model were acquired using a PGSE sequence with echo planar imaging (EPI) readout. A dense protocol of 46 $b$-values was employed, ranging from 0 to 4.5~ms/$\mu$m$^2$ in increments of 0.1~ms/$\mu$m$^2$. For each non-zero $b$-value, the diffusion encoding gradients were applied along three orthogonal directions. The timing parameters of diffusion encoding gradients were: $\delta = 24.1$~ms and $\Delta = 46.4$~ms. Other imaging parameters included: repetition time (TR) = 4000~ms, echo time (TE) = 101~ms, field of view (FOV) = $220 \times 220$~mm$^2$, matrix = $180 \times 180$, in-plane resolution = $1.2 \times 1.2$~mm$^2$, slice thickness = 5~mm. Total acquisition time was approximately 12~min for a volunteer. Standard clinical sequences, including T1WI, T2WI, T2-FLAIR, CE-T1WI, and standard DTI, were also acquired for glioma patients. Healthy subjects only underwent T1WI, T2WI, and DTI.

\subsection{Data processing}

dMRI images were preprocessed using the FMRIB Software Library (FSL, Oxford, UK)\cite{bib46}, including correcting for the artifacts from head motion, and the distortion from eddy current and susceptibility-induced field inhomogeneity, registering diffusion images to high-resolution T1WI images, and then resampling to an in-plane resolution of 1.2~mm. For each voxel, all dMRI signals were normalized to $S_0$. Voxel-wise $\mathcal{P}_D$ estimation was then performed using an in-house implementation of the IPDA algorithm in MATLAB (MathWorks, Natick, MA), utilizing dMRI signals measured from three orthogonal directions (Dir 1, Dir 2, Dir 3), and their direction-averaged values (Dir mean). Conventional dMRI parameters were calculated as follows: ADC was derived using two $b$-values (0, and 1.0~ms/$\mu$m$^2$), while MD and MK derived from kurtosis model were calculated using five $b$-values (0, 0.5, 1.0, 1.5, and 2.0~ms/$\mu$m$^2$).

To fully exploit the rich microstructural information inherent in $\mathcal{P}_D$, we defined thirteen quantitative parameters, categorized into global and local parameters. The global parameters capture the overall IDPD profile, including the mean ($\mu_{\mathrm{D}}$), standard deviation ($\sigma_{\mathrm{D}}$), skewness ($\operatorname{skew}_{\mathrm{D}}$), kurtosis ($\operatorname{kurt}_{\mathrm{D}}$), and the standard deviation of probability ($\sigma_{\mathrm{P}}$).

\paragraph{Mean diffusivity}
The $\mu_{\mathrm{D}}$ was defined as the first moment of $\mathcal{P}_D$:
\begin{equation}
\mu_{\mathrm{D}} = \sum_{n=1}^{N_D} p(D_n) \cdot D_n \label{eq:muD}
\end{equation}

\paragraph{Standard deviation of diffusivity}
The $\sigma_{\mathrm{D}}$ was defined as the square root of the second central moment of $\mathcal{P}_D$:
\begin{equation}
\sigma_{\mathrm{D}} = \sqrt{\sum_{n=1}^{N_D} p(D_n) \cdot (D_n - \mu_{\mathrm{D}})^2} \label{eq:sigmaD}
\end{equation}

\paragraph{Skewness}
The $\operatorname{skew}_{\mathrm{D}}$ was defined as the third standardized moment of $\mathcal{P}_D$:
\begin{equation}
\operatorname{skew}_{\mathrm{D}} = \sum_{n=1}^{N_D} p(D_n) \cdot \left(\frac{D_n - \mu_{\mathrm{D}}}{\sigma_{\mathrm{D}}}\right)^3 \label{eq:skewD}
\end{equation}

\paragraph{Kurtosis}
The $\operatorname{kurt}_{\mathrm{D}}$ was defined as the fourth standardized moment of $\mathcal{P}_D$:
\begin{equation}
\operatorname{kurt}_{\mathrm{D}} = \sum_{n=1}^{N_D} p(D_n) \cdot \left(\frac{D_n - \mu_{\mathrm{D}}}{\sigma_{\mathrm{D}}}\right)^4 \label{eq:kurtD}
\end{equation}

\paragraph{Standard deviation of probability}
The $\sigma_{\mathrm{P}}$ was defined as the standard deviation of probabilities:
\begin{equation}
\sigma_{\mathrm{P}} = \sqrt{\frac{1}{N_D} \sum_{n=1}^{N_D} \left(p(D_n) - \mu_{\mathrm{P}}\right)^2} \label{eq:sigmaP}
\end{equation}
where $\mu_{\mathrm{P}} = \frac{1}{N_D} \sum_{n=1}^{N_D} p(D_n)$ represents the mean probability in $\mathcal{P}_D$.

\paragraph{Local parameters}
The local parameters characterize distinct diffusivity regimes defined by diffusivity thresholds. In this study, three empirically determined thresholds ($D_{\mathrm{th},1} = 0.5$, $D_{\mathrm{th},2} = 1.5$, and $D_{\mathrm{th},3} = 3.0$), derived from the ROI-based analysis, were employed to partition the $\mathcal{P}_D$ into four distinct diffusivity regimes: low ($0 < D \leq D_{\mathrm{th},1}$), mid ($D_{\mathrm{th},1} < D \leq D_{\mathrm{th},2}$), high ($D_{\mathrm{th},2} < D \leq D_{\mathrm{th},3}$), and free ($D > D_{\mathrm{th},3}$). For each diffusivity regime, two local parameters were calculated, yielding a total of eight metrics: cumulative probabilities ($P_{\mathrm{low}}$, $P_{\mathrm{mid}}$, $P_{\mathrm{high}}$, and $P_{\mathrm{free}}$), which quantify the relative signal fraction of water molecules within each respective regime; mean diffusivities ($\mathrm{MD}_{\mathrm{low}}$, $\mathrm{MD}_{\mathrm{mid}}$, $\mathrm{MD}_{\mathrm{high}}$, and $\mathrm{MD}_{\mathrm{free}}$), which reflect the average diffusion rate within each regime.

Notably, the classification into these four diffusivity regimes serves merely as an illustrative example. Given the rich microstructural information inherent in $\mathcal{P}_D$, the partitioning can be customized according to underlying biophysical mechanisms, allowing for the isolation of specific diffusivity ranges that correspond to diverse physiological or pathological conditions.

\bibliography{sn-bibliography}% common bib file

\clearpage
\section*{Figures and Table}\label{sec6}

\begin{table}[htbp]
\centering
\caption{Pseudocode for IPDA algorithm}
\label{tab:ipda_pseudocode}
\renewcommand{\arraystretch}{1.8}
\begin{tabular}{@{}p{0.9\textwidth}@{}}
\toprule
\textbf{IPDA algorithm for quantifying IDPD} \\
\toprule
\textbf{Input:} dMRI signal, $S_0$ and $S(b_m)$ acquired at $b = 0$ ms/$\mu$m$^2$ and non-zero $b$-values $b_m$, $m \in \{1, 2, \ldots, M_b\}$ \\
\textbf{Output:} IDPD, $\mathcal{P}_D$ \\
\textbf{Initial Optimization:} \\
\hspace{1.5em}1. Calculate ADC$_m$ for each non-zero $b$-value via equation~(7); \\
\hspace{1.5em}2. Initialize $\pi_m$ as $1 / M_b$; \\
\hspace{1.5em}3. Initialize $\sigma_m$ via equations~(10) and~(11); \\
\hspace{1.5em}4. Initialize $M_b$ Gaussian components $\mathcal{N}_m(D \mid \text{ADC}_m, \sigma_m^2)$; \\
\hspace{1.5em}5. Initialize $f^{(0)}(D)$ as weighted mixture of $M_b$ Gaussian components via equation~(9); \\
\hspace{1.5em}6. Initialize diffusivity limits $D_{\min}^{(0)}$ and $D_{\max}^{(0)}$ via equation~(12), and number of intervals $N_D^{(0)}$; \\
\hspace{1.5em}7. Initialize $\mathcal{P}_{\text{prior}}^{(0)}$ by integrating $f^{(0)}(D)$ across $N_D^{(0)}$ intervals via equation~(13); \\
\hspace{1.5em}8. Initialize probability bounds via equation~(14); \\
\hspace{1.5em}9. Estimate $\mathcal{P}_D^{(0)}$ by regularized NNLS estimation via equation~(15). \\
\textbf{Iterative Optimization:} \\
\hspace{1.5em}\textbf{for $k = 1$ to $K$ do} \\
\hspace{3em}1. Update $\pi_n^{(k)} = p^{(k-1)}(D_n)$, $n \in \{1, 2, \ldots, N_D^{(k-1)}\}$; \\
\hspace{3em}2. Update $\sigma_n^{(k)}$ based on equations~(10) and~(11); \\
\hspace{3em}3. Update $N_D^{(k-1)}$ Gaussian components $\mathcal{N}_n(D \mid D_n, \sigma_n^2)$; \\
\hspace{3em}4. Update $f^{(k)}(D)$ as weighted mixture of $N_D^{(k-1)}$ Gaussian components via equation~(16); \\
\hspace{3em}5. Update diffusivity limits $D_{\min}^{(k)}$ and $D_{\max}^{(k)}$ via equation~(18), and number of intervals $N_D^{(k)}$; \\
\hspace{3em}6. Update $\mathcal{P}_{\text{prior}}^{(k)}$ by integrating $f^{(k)}(D)$ across $N_D^{(k)}$ intervals based on equation~(13); \\
\hspace{3em}7. Update probability bounds based on equation~(14); \\
\hspace{3em}8. Estimate $\mathcal{P}_D^{(k)}$ by regularized NNLS estimation based on equation~(15); \\
\hspace{3em}9. Termination condition: if $\sum_{n=1}^{N_D^{(k)}} \left| p^{(k)}(D_n) - p^{(k-1)}(D_n) \right| < 1 \times 10^{-3}$ or $k = 10$ then break; \\
\hspace{1.5em}\textbf{end for} \\
\textbf{Return:} $\mathcal{P}_D \leftarrow \mathcal{P}_D^{(k)}$ \\
\bottomrule
\end{tabular}
\end{table}

% figures
% ========== Figure 1 ==========
\begin{figure}[htbp]
\centering
\includegraphics[width=\textwidth]{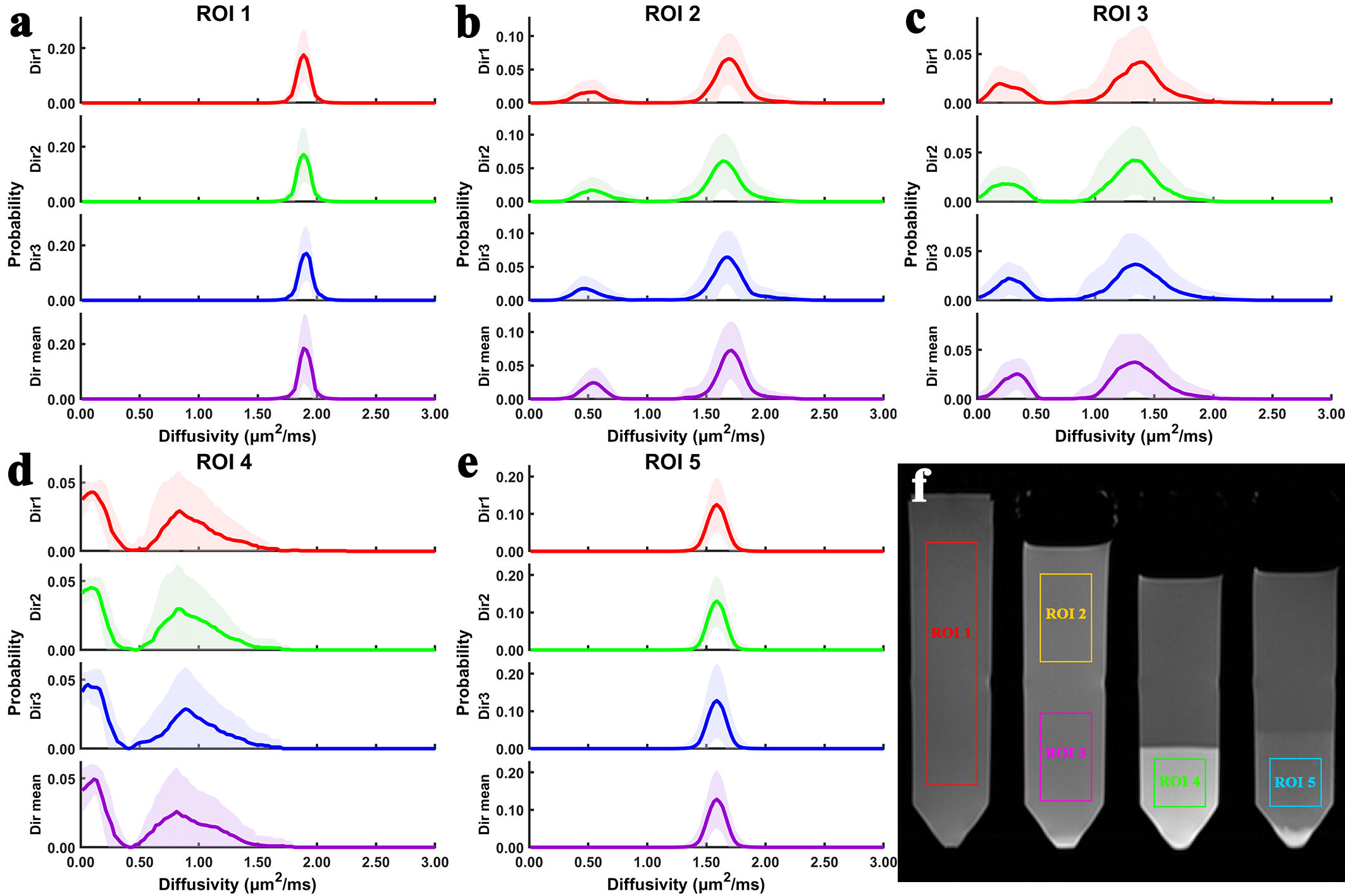}
\caption{ROI-based IDPD analysis in aqueous phantoms.
a--e, 2D plots showing the mean IDPDs across all voxels within five ROIs. These ROIs include a water phantom (ROI~1, a), the upper and lower layers of non-centrifuged yeast suspension (ROI~2 and 3, b and c), the lower layer of centrifuged yeast suspension (ROI~4, d), and the lower layer of yeast suspension following cell disruption and centrifugation (ROI~5, e). Each panel presents the mean probabilities (y-axis) corresponding to discrete diffusivities (x-axis) from three orthogonal diffusion encoding directions (Dir~1 [1,0,0]; Dir~2 [0,-1,0]; Dir~3 [0,0,1]) and their direction-averaged, while the colored shaded regions indicate the standard deviations of the probabilities across voxels.
f, T1-weighted image (T1WI) with color-coded overlays illustrating the spatial position of five ROIs.}
\label{fig:1}
\end{figure}

% ========== Figure 2 (跨页标题) ==========
\begin{figure}[htbp]
\centering
\includegraphics[height=\textheight, keepaspectratio]{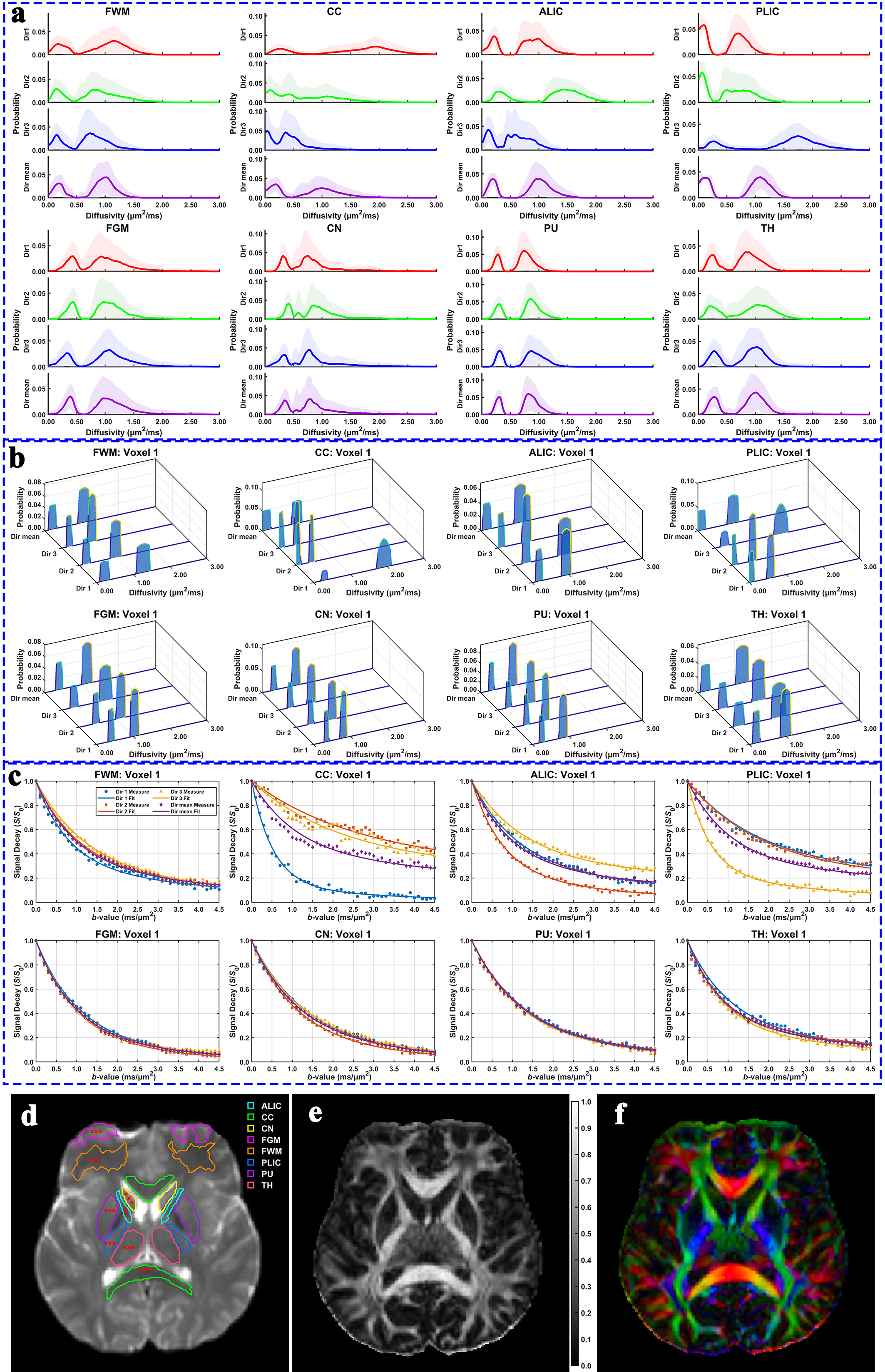}  % 请替换为您的图片文件名
\label{fig:2-img}  % 仅占位，实际引用使用 afterpage 内的 label
\end{figure}

\afterpage{%
\vspace*{-\topskip}% 消除页面顶部默认空白，使内容从第1行开始
\refstepcounter{figure}% 递增 figure 计数器
\phantomsection
\addcontentsline{toc}{subsection}{\figurename~\thefigure}% 大纲条目
\noindent\textbf{\figurename~\thefigure: }%
ROI-based and voxel-based IDPD analysis in a healthy human brain.
\textbf{a,} 2D plots showing the mean IDPDs across all voxels within eight ROIs. These ROIs include four WM ROIs (FWM, CC, ALIC, PLIC) and four GM ROIs (FGM, PU, CN, TH). Each panel presents the mean probabilities (y-axis) corresponding to discrete diffusivities (x-axis) from three orthogonal diffusion encoding directions (Dir~1 [1,0,0]; Dir~2 [0,-1,0]; Dir~3 [0,0,1]) and their direction-averaged, while the colored shaded regions indicate the standard deviations of the probabilities across voxels.\textbf{b,} 3D waterfall plots showing the IDPDs of representative voxels. Only the first voxel selected from each of the eight ROIs is displayed here. The IDPDs for the remaining 16 voxels (two additional voxels per ROI) are provided in Extended Data Fig.~\ref{fig:ed1}a. The x-axis represents diffusivity, the y-axis represents direction, and the z-axis represents probability.\textbf{c,} 2D plots showing the measured and fitted signal decays of eight representative voxels. Results for the other 16 voxels across the eight ROIs are shown in Extended Data Fig.~\ref{fig:ed1}b. The colored scatter points and solid lines respectively represent the measured and fitted signal decays (y-axis) at varying b-values (x-axis, from 0 to 4.5~ms/$\mu$m$^2$) along different directions.\textbf{d,} Anatomical locations of the analyzed ROIs and voxels. The $b_0$ image serves as an anatomical reference. The colored contours delineate the boundaries of eight ROIs, where `x' symbols mark the precise spatial locations of twenty-four voxels. Three voxels were selected from each ROI.\textbf{e,} Grayscale fractional anisotropy (FA) map derived from tensor model providing a reference for diffusion anisotropy.\textbf{f,} Color-coded FA map illustrating the principal fiber orientations (red: left-right; green: anterior-posterior; blue: superior-inferior).

\label{fig:2}
}

% ========== Figure 3 (与 Figure 2 完全相同的方式) ==========
\clearpage
\begin{figure}[htbp]
\centering
\includegraphics[height=\textheight, keepaspectratio]{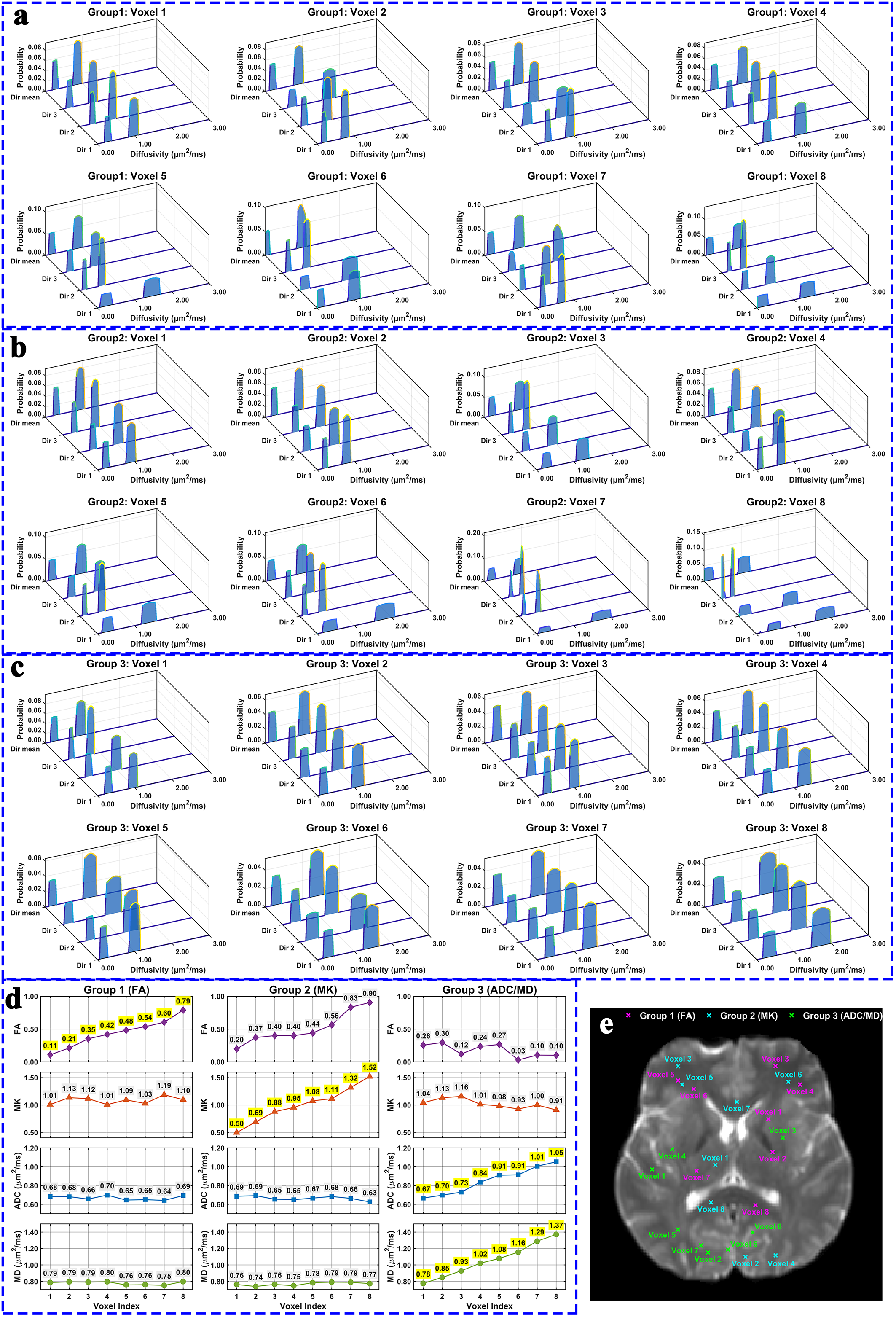}
\label{fig:3-img}
\end{figure}

\afterpage{%
\vspace*{-\topskip}% 从第1行开始
\refstepcounter{figure}% 递增 figure 计数器
\phantomsection
\addcontentsline{toc}{subsection}{\figurename~\thefigure}% 大纲条目
\noindent\textbf{\figurename~\thefigure: }%
Voxel-based comparative analysis of IDPD and conventional dMRI parameter in a healthy human brain. Representative voxels ($n=24$) were categorized into three groups ($n=8$, per group) to illustrate the IDPD evolution in response to variations in conventional dMRI parameters.
\textbf{a,} IDPD evolution with increasing FA. 3D waterfall plots of eight voxels, selected across different brain regions and sorted by ascending FA values, highlight the transition of the diffusivity distribution as macroscopic anisotropy increases.
\textbf{b,} IDPD evolution with increasing MK. 3D waterfall plots of eight voxels sorted by ascending MK values demonstrate how the diffusivity distribution broadens or shifts in response to increasing microstructural complexity.
\textbf{c,} IDPD evolution with increasing ADC and MD values. 3D waterfall plots of eight voxels sorted by ascending ADC and MD values illustrate the shift in diffusivity distribution associated with changes in the overall magnitude of water mobility.
\textbf{d,} Evolution of conventional dMRI parameters for the selected twenty-four voxels. 2D line plots display the specific values of FA, MK, and ADC/MD for each of the eight voxels within their respective groups.
\textbf{e,} Anatomical locations of the analyzed voxels. The $b_0$ image serves as an anatomical reference, where `x' symbols mark the precise spatial locations of the twenty-four voxels. The symbols are color-coded to distinguish between the three groups.

\label{fig:3}%
}

% ========== Figure 4 ==========
\clearpage
\begin{figure}[htbp]
\centering
\includegraphics[width=\textwidth]{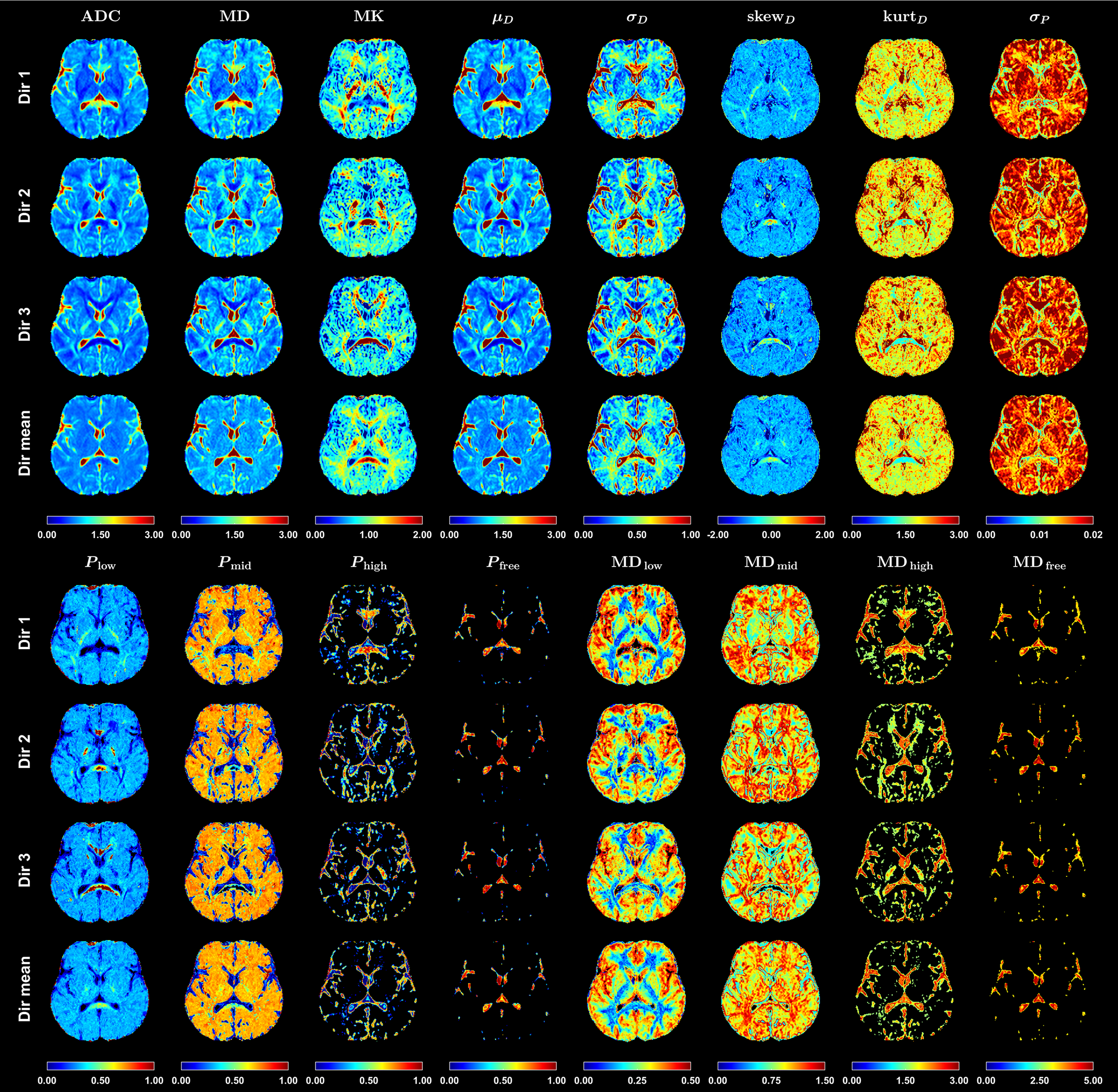}
\phantomsection
\addcontentsline{toc}{subsection}{\figurename~\thefigure}
\caption{Representative parametric maps of a healthy volunteer. Sixteen parameters include ADC, MD, MK, and thirteen IDPD-derived parameters. Except for ADC, MD, $\mu_{\text{D}}$, $\sigma_{\text{D}}$, MD$_{\text{low}}$, MD$_{\text{mid}}$, MD$_{\text{high}}$, and MD$_{\text{free}}$ (measured in $\mu$m$^2$/ms), all other metrics are dimensionless. Color bars indicate the range for each parameter. Rows correspond to maps along different directions (Dir~1: [1,0,0]; Dir~2: [0,-1,0]; Dir~3: [0,0,1]; and Dir mean).}
\label{fig:4}
\end{figure}

% ========== Figure 5 ==========
\clearpage
\begin{figure}[htbp]
\centering
\includegraphics[width=\textwidth]{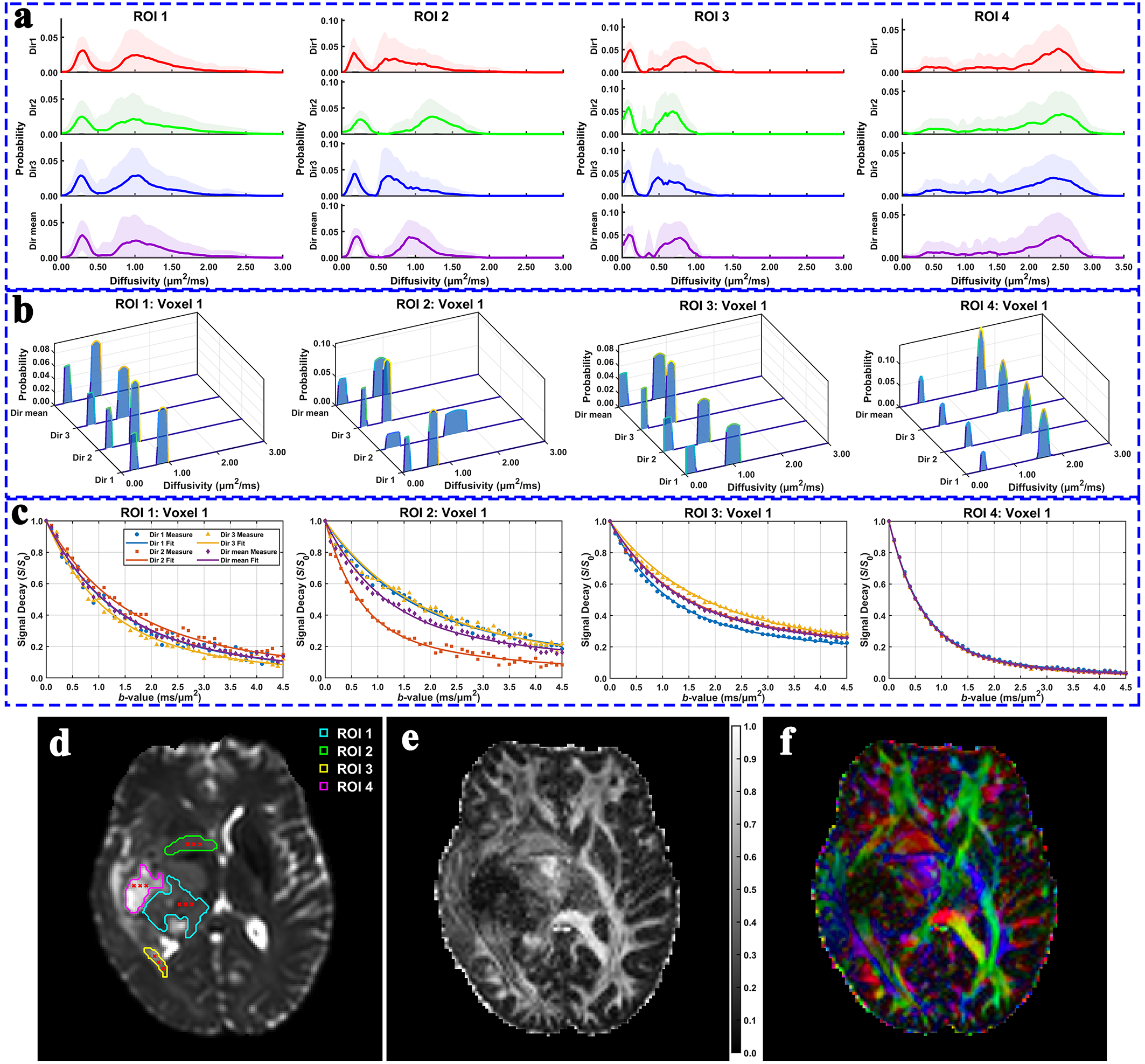}
\label{fig:5-img}  % 占位标签（实际引用用下面的 label）
\end{figure}

% 手动输出标题（允许跨页）
\refstepcounter{figure}%
\phantomsection
\addcontentsline{toc}{subsection}{\figurename~\thefigure}%
\noindent\textbf{\figurename~\thefigure: }%
ROI-based and voxel-based IDPD analysis in brain of a high-grade glioma patient (WHO grade IV glioblastoma).
\textbf{a,} 2D plots showing the mean IDPDs across all voxels within four ROIs. These ROIs are defined as follows: ROI~1 denotes the enhancement area on CE-T1WI, representing active solid tumor; ROI~2 and ROI~3 correspond to tumor infiltration within the right association fibers and optic radiation, respectively; and ROI~4 targets the non-enhancing core on CE-T1WI, indicating central necrosis. Each panel presents the mean probabilities (y-axis) corresponding to discrete diffusivities (x-axis) from three orthogonal diffusion encoding directions (Dir~1: [0.656,0.402,0.636]; Dir~2: [0.712,-0.611,-0.343]; Dir~3: [0.250,0.682,-0.692]) and their direction-averaged, while the colored shaded regions indicate the standard deviations of the probabilities across voxels.
\textbf{b,} 3D waterfall plots showing the IDPDs of four representative voxels. Only the first voxel selected from each of the four ROIs is displayed here. The IDPDs for the remaining eight voxels (two additional voxels per ROI) are provided in Extended Data Fig.~\ref{fig:ed4}a. The x-axis represents diffusivity, the y-axis represents direction, and the z-axis represents probability.
\textbf{c,} 2D plots showing the measured and fitted signal decays of four representative voxels. Results for the other 16 voxels across the eight ROIs are shown in Extended Data Fig.~\ref{fig:ed4}b. The colored scatter points and solid lines respectively represent the measured and fitted signal decays (y-axis) at varying b-values (x-axis, from 0 to 4.5~ms/$\mu$m$^2$) along different directions.
\textbf{d,} Anatomical locations of the analyzed ROIs and voxels. The $b_0$ image serves as an anatomical reference. The colored contours delineate the boundaries of four ROIs, where `x' symbols mark the precise spatial locations of twenty-four voxels. Three voxels were selected from each ROI.
\textbf{e,f,} WM tracts integrity. Grayscale (e) and color-coded (f) FA maps illustrating varying degrees of disruption or displacement of WM tracts in the right hemisphere.

\label{fig:5}

% ========== Figure 6 ==========
\clearpage
\begin{figure}[htbp]
\centering
\includegraphics[width=\textwidth]{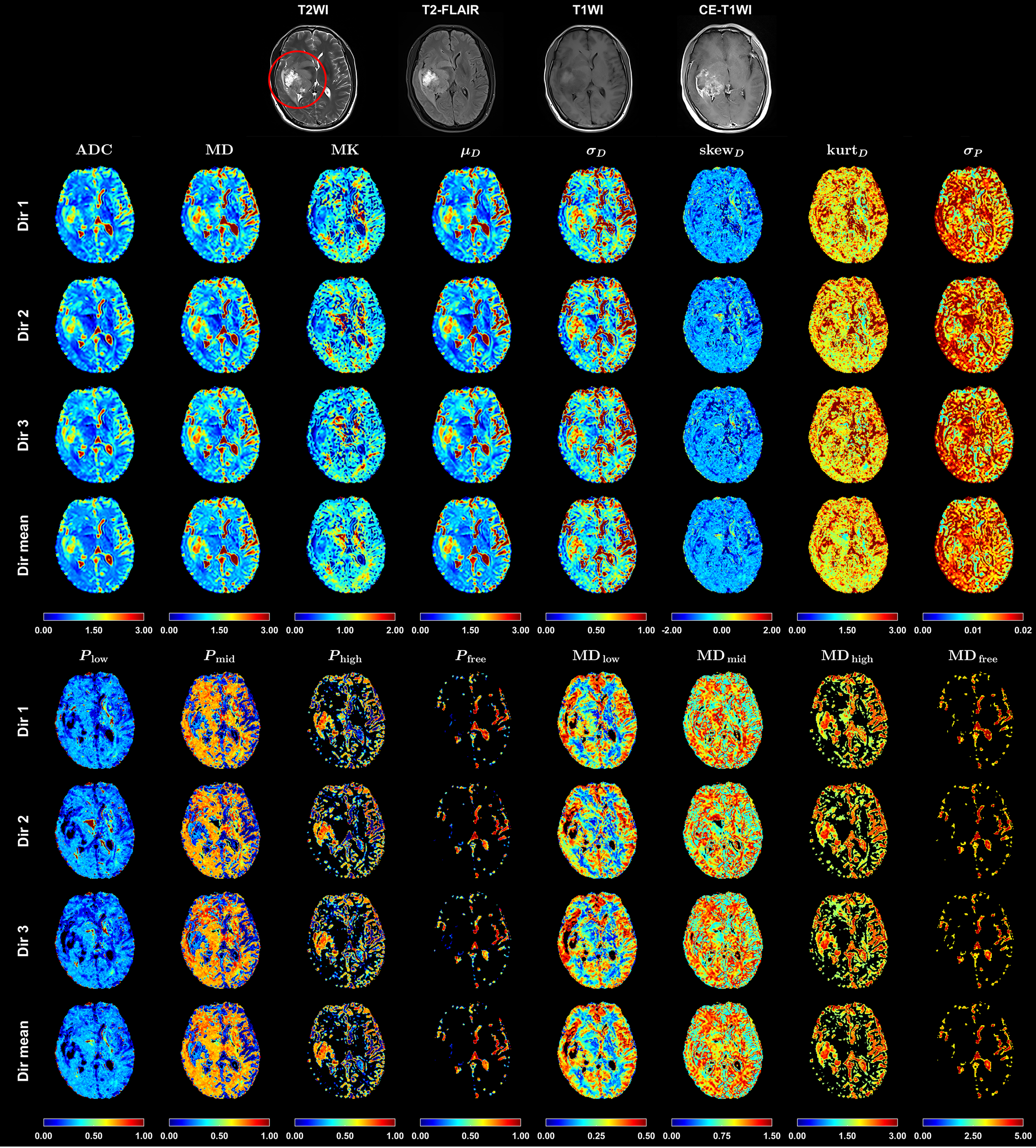}
\label{fig:6-img}  % 占位标签（实际引用用下面的 label）
\end{figure}

% 手动输出标题（允许跨页）
\refstepcounter{figure}%
\phantomsection
\addcontentsline{toc}{subsection}{\figurename~\thefigure}%
\noindent\textbf{\figurename~\thefigure: }%
Routine MRI images and representative parametric maps of a male high-grade glioma patient (WHO grade IV glioblastoma). Routine MRI images (top row), including T2-weighted imaging (T2WI), T2-fluid attenuated inversion recovery (T2-FLAIR), T1WI, and contrast-enhanced T1WI (CE-T1WI), reveal a large irregular, infiltrative lesion extending from the right temporal lobe to the right basal ganglia, thalamus, and corpus callosum (marked with red circle on T2WI image). The lesion exhibits a typical infiltrative growth pattern, crossing the midline to involve the contralateral side, which is a characteristic feature of high-grade glioma spreading along white matter tracts. The lesion area shows mixed hyperintensity on both T2WI and T2-FLAIR. The central hyperintensity typically represents necrosis, whereas the hyperintensity at the lesion margin reflects a combination of tumor infiltration and vasogenic edema. Pre-contrast T1WI demonstrates mixed hypo- to iso-intensity. Post-contrast CE-T1WI reveals irregular, thick-walled ``garland-like'' enhancement. The central non-enhancing area suggests necrosis. Parametric maps, including ADC, MD, MK and thirteen IDPD metrics, revealing the microstructural alterations within the lesion areas. Except for ADC, MD, $\mu_{\text{D}}$, $\sigma_{\text{D}}$, MD$_{\text{low}}$, MD$_{\text{mid}}$, MD$_{\text{high}}$, and MD$_{\text{free}}$ (measured in $\mu$m$^2$/ms), all other metrics are dimensionless. Color bars indicate the range for each parameter. Rows correspond to maps along different directions (Dir~1: [0.656,0.402,0.636]; Dir~2: [0.712,-0.611,-0.343]; Dir~3: [0.250, 0.682,-0.692]; and Dir mean). These diffusion encoding directions are distinct from the schemes employed in healthy human brain imaging, aiming to explore microstructural properties across a more diverse spatial distribution.

\label{fig:6}

% ========== Extended Data Figures Section ==========
\section*{Extended Data Figures}
\addcontentsline{toc}{section}{Extended Data Figures}

% ========== Extended Data Fig. 1 ==========
\begin{figure}[htbp]
\centering
\includegraphics[height=\textheight, keepaspectratio]{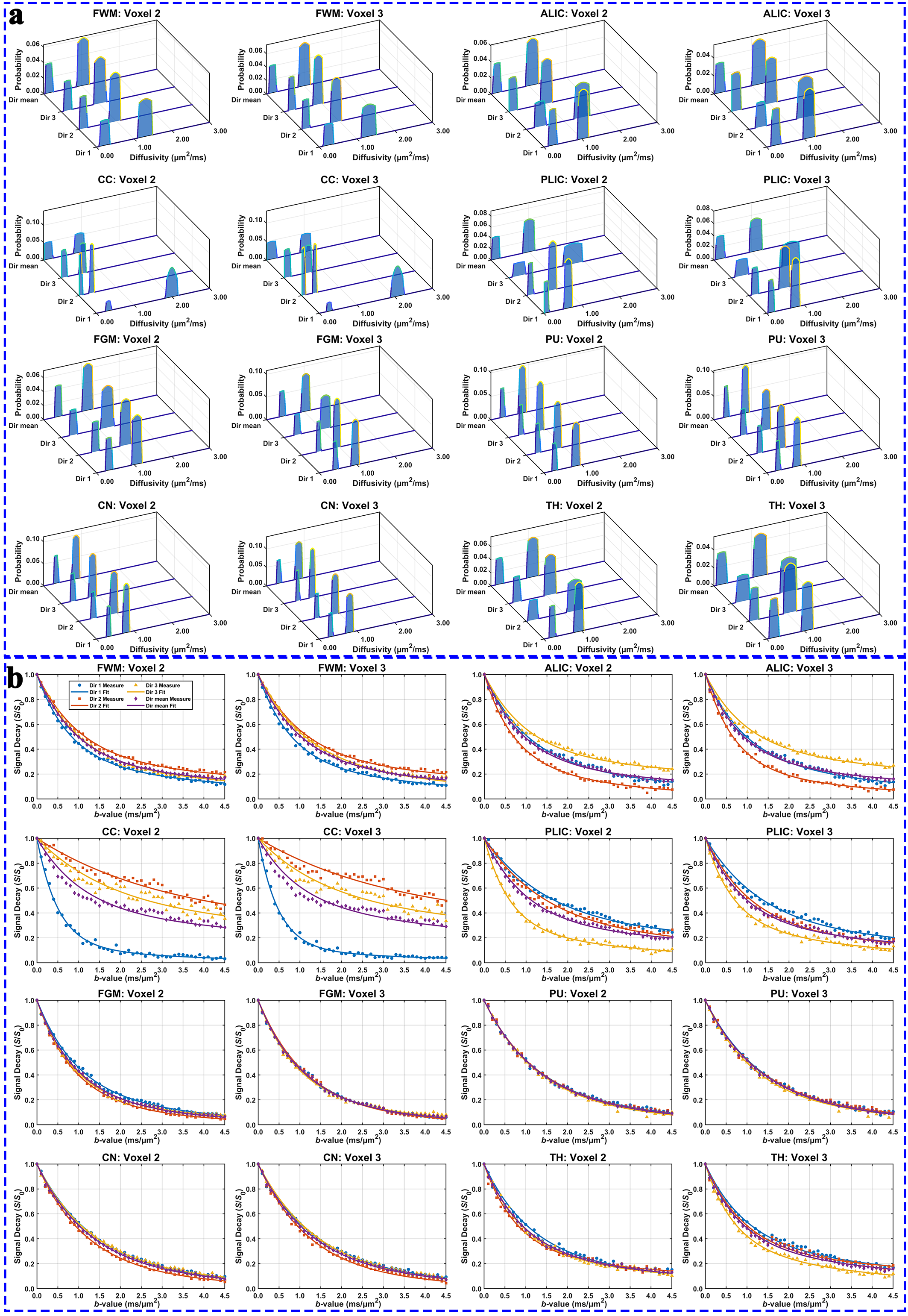}
\label{fig:ed1-img}
\end{figure}

\afterpage{%
\vspace*{-\topskip}
\refstepcounter{edfig}% 使用独立计数器
\phantomsection
\addcontentsline{toc}{subsection}{\theedfig}
\noindent\textbf{\theedfig: }%
Supplemental voxel-based IDPD analysis for the remaining sixteen voxels described in Fig.~\ref{fig:2}.
\textbf{a,} 3D waterfall plots showing the IDPDs of the remaining sixteen voxels. These plots correspond to the second and third voxels selected from each of the eight ROIs. The x-axis represents diffusivity, the y-axis represents direction, and the z-axis represents probability.
\textbf{b,} 2D plots showing the measured and fitted signal decays of the remaining sixteen voxels. The colored scatter points and solid lines respectively represent the measured and fitted signal decays (y-axis) at varying b-values (x-axis, from 0 to 4.5~ms/$\mu$m$^2$) along different directions.

\label{fig:ed1}
}
\clearpage

% ========== Extended Data Fig. 2 ==========
\clearpage
\begin{figure}[htbp]
\centering
\includegraphics[width=\textwidth]{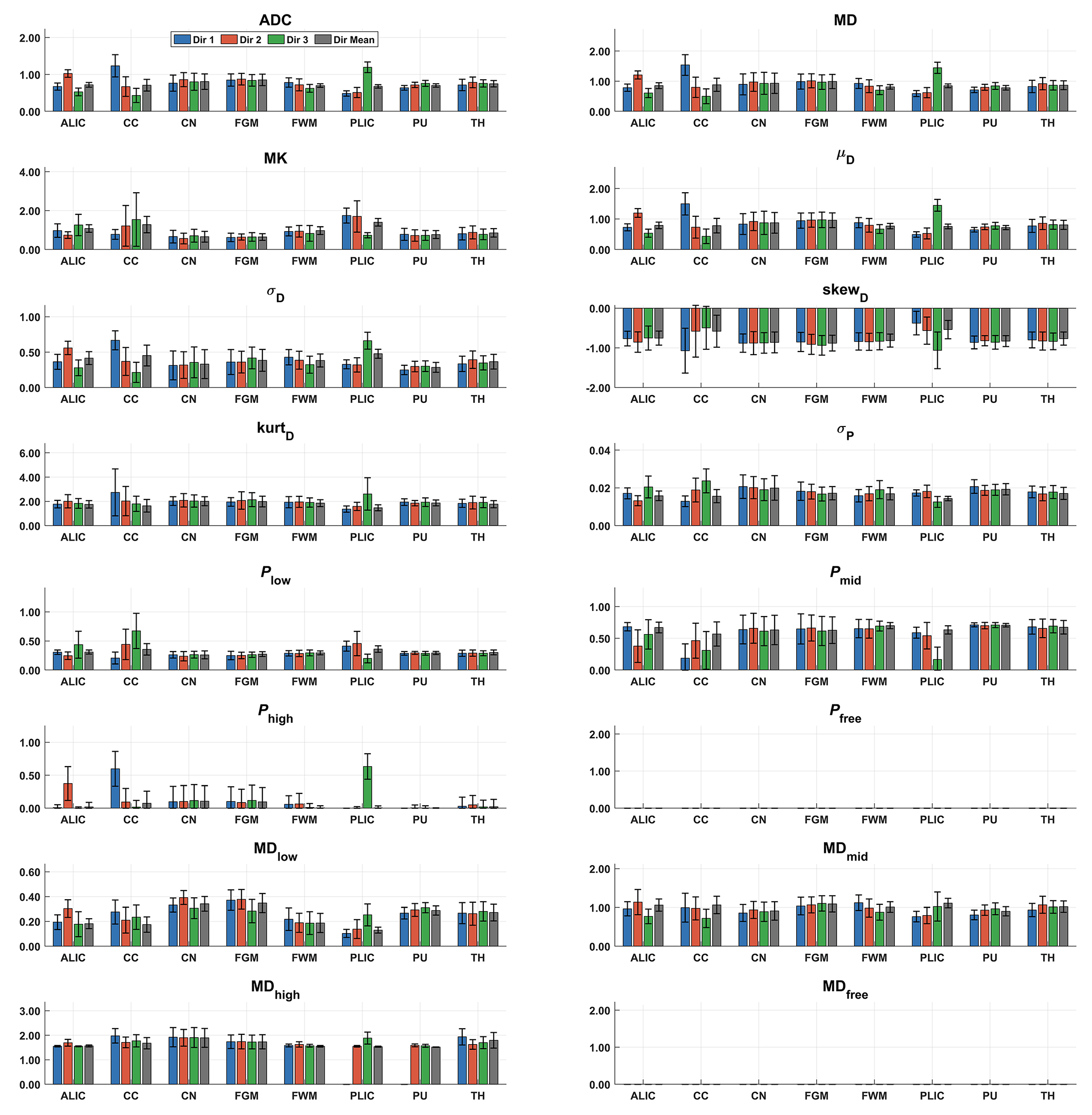}
\label{fig:ed2-img}
\end{figure}

\vspace*{-\topskip}
\refstepcounter{edfig}%
\phantomsection
\addcontentsline{toc}{subsection}{\theedfig}
\noindent\textbf{\theedfig: }%
Regional comparison and directional dependence of conventional and IDPD-derived parameters in healthy human brain. Each panel represents one of sixteen parameters, including ADC, MD, MK, and thirteen IDPD-derived parameters. The bar charts illustrate the means and standard deviations of each parameter across eight ROIs (ROI~1--4) and four directions (Dir~1: [1,0,0]; Dir~2: [0,-1,0]; Dir~3: [0,0,1]; and Dir mean).

\label{fig:ed2}

% ========== Extended Data Fig. 3 ==========
\clearpage
\begin{figure}[htbp]
\centering
\includegraphics[width=\textwidth]{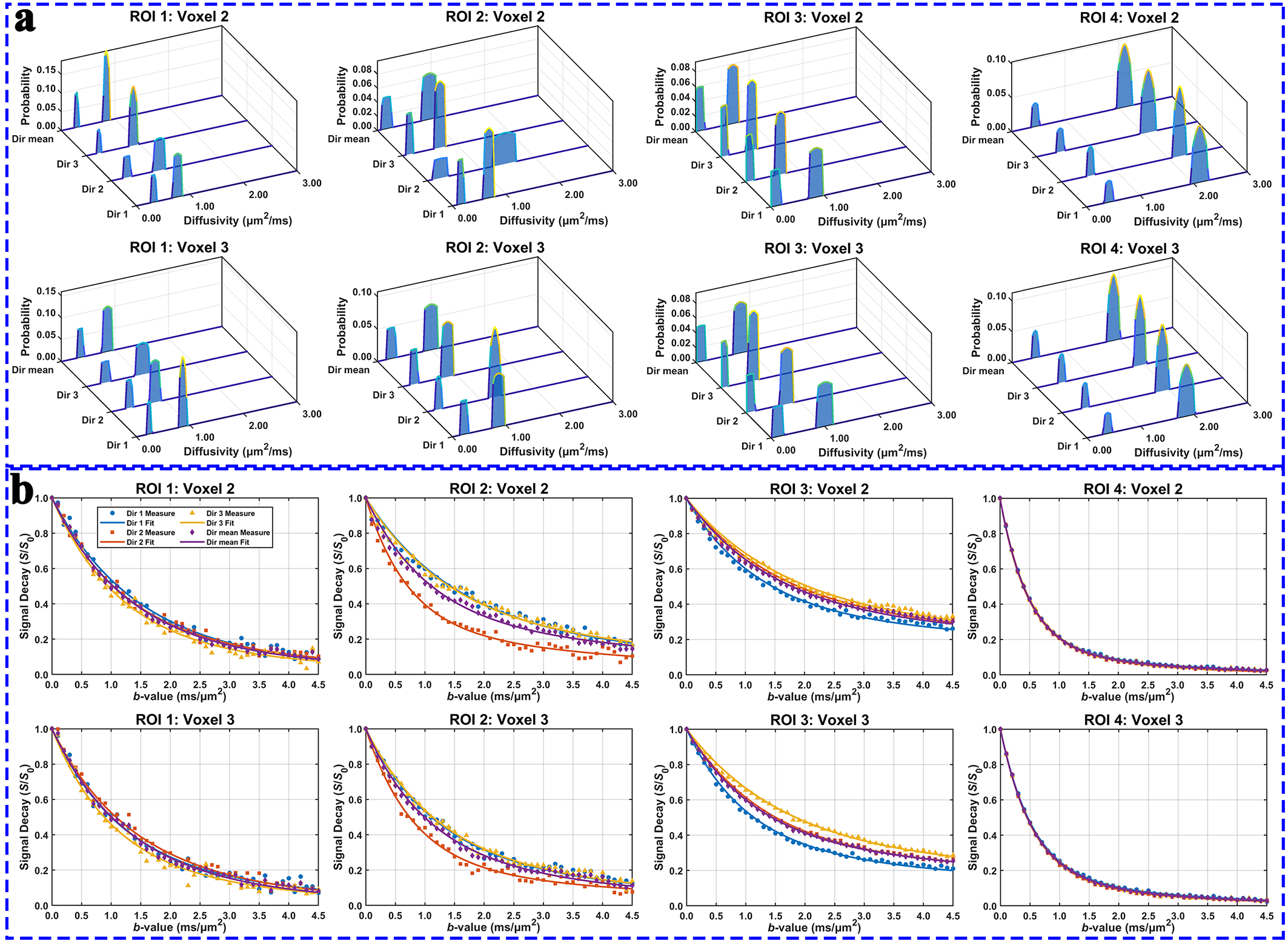}
\label{fig:ed3-img}
\end{figure}

\refstepcounter{edfig}%
\phantomsection
\addcontentsline{toc}{subsection}{\theedfig}
\noindent\textbf{\theedfig: }%
Supplemental voxel-based IDPD analysis for the remaining eight voxels described in Fig.~\ref{fig:5}.
\textbf{a,} 3D waterfall plots showing the IDPDs of the remaining eight voxels. These plots correspond to the second and third voxels selected from each of the four ROIs. The x-axis represents diffusivity, the y-axis represents direction, and the z-axis represents probability.
\textbf{b,} 2D plots showing the measured and fitted signal decays of the remaining eight voxels. The colored scatter points and solid lines respectively represent the measured and fitted signal decays (y-axis) at varying b-values (x-axis, from 0 to 4.5~ms/$\mu$m$^2$) along different directions.

\label{fig:ed3}

% ========== Extended Data Fig. 4 ==========
\clearpage
\begin{figure}[htbp]
\centering
\includegraphics[width=\textwidth]{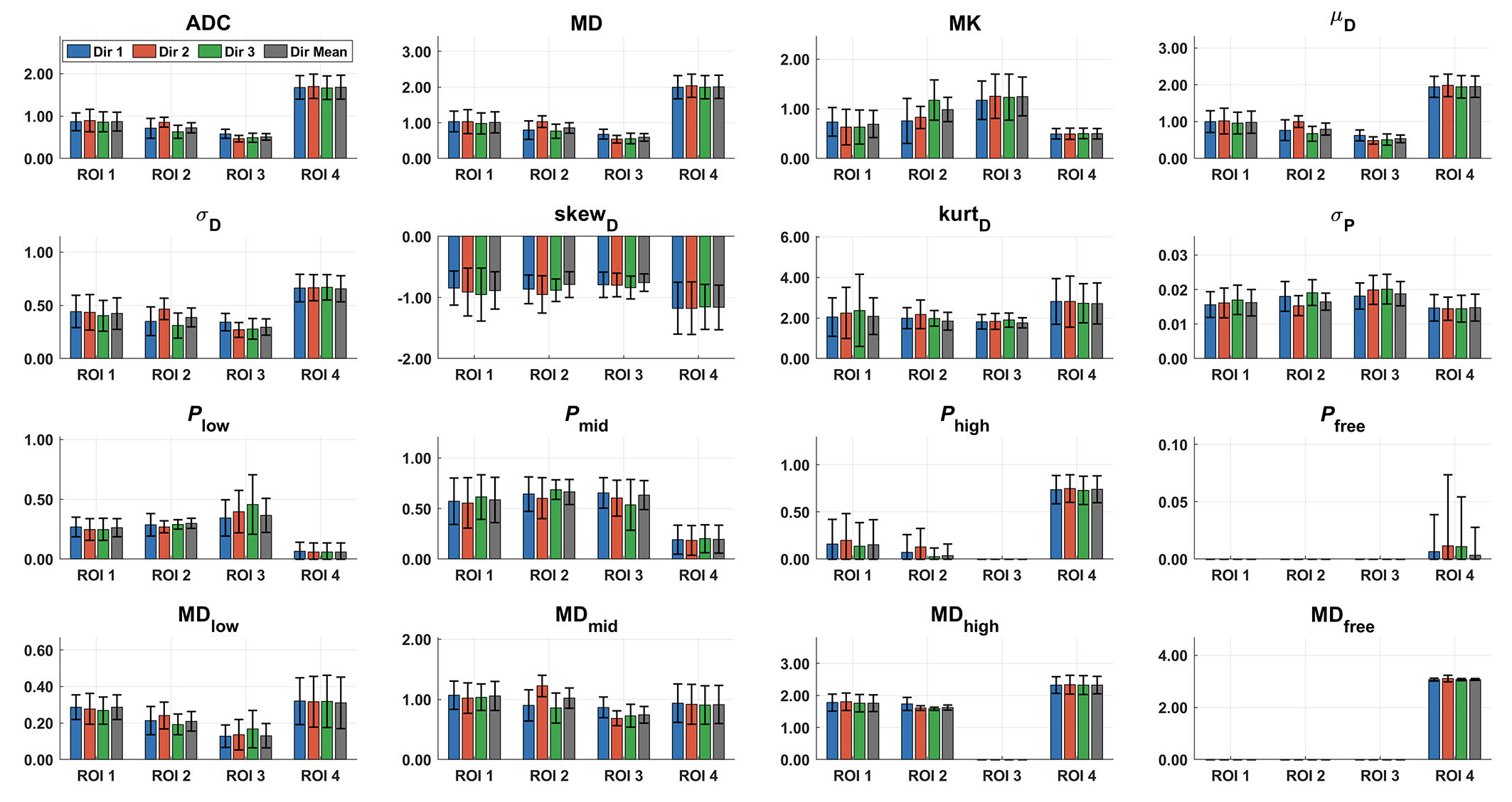}
\label{fig:ed4-img}
\end{figure}

\refstepcounter{edfig}%
\phantomsection
\addcontentsline{toc}{subsection}{\theedfig}
\noindent\textbf{\theedfig: }%
Regional comparison and directional dependence of conventional and IDPD-derived parameters of a high-grade glioma patient. Each panel represents one of sixteen parameters, including ADC, MD, MK, and thirteen IDPD-derived parameters. The bar charts illustrate the means and standard deviations of each parameter across four ROIs (ROI~1--4) and four directions (Dir~1: [0.656, 0.402, 0.636]; Dir~2: [0.712, -0.611, -0.343]; Dir~3: [0.250, 0.682, -0.692]; and Dir mean).

\label{fig:ed4}

\end{document}